%% file: NPJ-Aliasing-NF.tex
\documentclass[journal,10pt]{IEEEtran}

\input{heading/packages}

\IEEEoverridecommandlockouts

\title{Aliasing in Near-Field Array Ambiguity Functions: a Spatial Frequency-Domain Framework}

\author{\IEEEauthorblockN{
Gilles Monnoyer$^\dagger$, Jérôme Louveaux, Laurence Defraigne, Baptiste Sambon, Luc Vandendorpe}

\IEEEauthorblockA{\textit{ICTEAM, UCLouvain, Louvain-la-Neuve (Belgium), emails~: firstname.lastname@uclouvain.be \\
{\small $^\dagger$ corresponding author email address: gilles.monnoyer@uclouvain.be}
\vspace{-5mm}} 
}
}

\begin{document}
\input{heading/shortcut}

\input{heading/notations}
\input{heading/acronyms}

\bstctlcite{MyBSTcontrol}

\maketitle

\begin{abstract} 
    Next-generation communication and localization systems increasingly rely on \glspl{xla}, which promise unprecedented spatial resolution and new functionalities.
    These gains arise from their inherent operation in the \gls{nf} regime, where the spherical nature of the wavefront can no longer be ignored. Consequently, characterizing the \acrlong{af}--which amounts to the matched beampattern-- is considerably more challenging.
    Implementing \acrlongpl{xla} with half-wavelength element spacing is costly and complex, and removing antennas introduces intricate aliasing structures, \ie grating lobes.
    Whereas prior work has addressed \rev{the challenging modelling of grating lobes} using approximations tailored to specific array geometries, this paper develops a general framework that reveals their fundamental origins and geometric behavior in \gls{nf} \acrlongpl{af}.
    Using a local spatial-frequency analysis of steering signals, we derive a systematic methodology to model \gls{nf} grating lobes as aliasing artifacts, quantify their structure on the \acrlong{af}, and \rev{provide guidelines to design} \glspl{xla} operating within aliasing-safe regions. 
    We further connect \rev{the presented framework} to established \acrlong{ff} principles.
    Finally, we demonstrate the practical value of the approach by deriving closed-form expressions for aliasing-free regions in canonical \acrlongpl{ula} and \acrlongpl{uca}.
\end{abstract}

\glsresetall

\vspace{-0mm}

\input{sections/1-Intro}

\input{sections/2-Model}
\input{sections/3-Aliasing}

\input{sections/4-Chirp}
\input{sections/5-ULA}

\input{sections/6-UCA}

\input{sections/7-Conclusions}
\input{sections/8-Misc}
\input{sections/9-Appendix}

\bibliographystyle{IEEEtran}
\bibliography{nourl, references-NF-SL}

\end{document}

%% file: heading/packages.tex
\usepackage{amsmath,amssymb,amsfonts,amsthm,mathrsfs,mathtools,bbm,stmaryrd}
\SetSymbolFont{stmry}{bold}{U}{stmry}{m}{n}
\usepackage{xparse,pgffor,xspace,glossaries,url}
\usepackage[caption=false,font=normalsize,labelfont=sf,textfont=sf]{subfig}
\usepackage{xcolor}
\usepackage{cite}
\usepackage{graphicx}
\usepackage{ifthen}
\usepackage{svg}
\usepackage[hidelinks]{hyperref}
\usepackage{enumitem}
\usepackage[ruled,vlined]{algorithm2e}
\usepackage[normalem]{ulem}

\usepackage{footnote}
\makesavenoteenv{tabular}
\makesavenoteenv{table*}

\newtheorem{definition}{Definition}

\newtheorem{property}{Property}

\newtheorem{lemma}{Lemma}
\newtheorem{theorem}{Theorem}
\newtheorem{corollary}{Corollary}

%% file: heading/shortcut.tex
\definecolor{myBlue}{RGB}{30, 70, 200}

\newcommand{\cc}{\centering}
\newcommand{\bs}{\boldsymbol}
\newcommand{\fs}{\mathsf}
\newcommand{\bb}{\mathbb}
\newcommand{\cl}{\mathcal}
\newcommand{\bcl}[1]{\boldsymbol{\mathcal{#1}}}

\newcommand{\ts}{\textstyle}
\newcommand{\ie}{\emph{i.e.}, }
\newcommand{\eg}{\emph{e.g.}, }
\newcommand{\etal}{\emph{et al.}}
\newcommand{\st}{\ensuremath{\mathrm{s.t.}}\xspace}

\newcommand{\expe}{\mathrm e}
\newcommand{\conv}{\otimes}
\newcommand{\sign}{\mathrm{sign}}

\newcommand{\argmax}{\arg\,\max}

\newcommand{\todolist}[1]{
{\color{olive}{\bf{Todo List : \\} \small{\bf{#1}}}}
}

\newcommand{\GM}[1]{
{\color{myBlue} [ {\textbf GM : } {\small #1} ]}
}

\newcommand{\BS}[1]{
{\color{orange} [ {\textbf BS : } {\small #1} ]}
}

\newcommand{\LD}[1]{
{\color{purple} [ {\textbf LD : } {\small #1} ]}
}

\newcommand{\LV}[1]{
{\color{red} [ {\textbf LV : } {\small #1} ]}
}

\newcommand{\rev}[1]{#1}

\newcommand{\del}[1]{
    \textcolor{red}{\sout{#1}}
}

%% file: heading/notations.tex
\newcommand{\unitvec}{\hat{\bs e}}
\newcommand{\anglecasual}{\phi}
\newcommand{\indicator}{\eta}
\newcommand{\bessel}{\mathrm{J}}
\newcommand{\sinc}{\mathrm{sinc}}

\newcommand{\cel}{c}

\newcommand{\mtest}[1]{{\tilde{#1}}}
\newcommand{\mapprox}[1]{{\hat{#1}}}
\newcommand{\uca}{{\mathrm{uca}}}
\newcommand{\ula}{{\mathrm{ula}}}
\newcommand{\sampled}{\mathrm{sampled}}
\newcommand{\safe}{\mathrm{safe}}
\newcommand{\ff}{\mathrm{FF}}
\newcommand{\mtoy}{\mathrm{toy}}

\newcommand{\ambifun}{A}
\newcommand{\ambifunband}{\cl A}
\newcommand{\sambifun}{A_\mathrm{S}}
\newcommand{\diffaf}{D}
\newcommand{\matchspectrum}{G}
\newcommand{\smatchspectrum}{G_\mathrm{S}}
\newcommand{\matchspectrumphi}{G_\xi}
\newcommand{\matchspectrumabs}{G_\alpha}
\newcommand{\dirac}{\delta}
\newcommand{\freqvar}{\bs \omega}
\newcommand{\paramtf}{\nu}

\newcommand{\threshband}{\epsilon}

\newcommand{\antgrid}{\antdomain_\mathrm{G}}
\newcommand{\paramgrid}{\paramdomain_\mathrm{G}}

\newcommand{\sigtx}{s}
\newcommand{\sigrx}{s_\mathrm{R}}
\newcommand{\meas}{y}
\newcommand{\energy}{\mathcal E}
\newcommand{\sig}{h}
\newcommand{\phasefun}{\phi}
\newcommand{\amplitudefun}{a}
\newcommand{\matchfun}{g}
\newcommand{\smatchfun}{g_\mathrm{S}}
\newcommand{\matchamp}{\alpha}
\newcommand{\matchphase}{\xi}

\newcommand{\bandlim}{K}
\newcommand{\bandlimplus}{B_{1}}
\newcommand{\bandlimminus}{B_{-1}}
\newcommand{\bandlimpm}{B_{q}}
\newcommand{\bandlimsec}{K^{\mathrm{s}}}
\newcommand{\kset}{\mathcal K}
\newcommand{\bandlimstrict}{\bar K_\mathrm{0}}
\newcommand{\afr}{\mathcal S}
\newcommand{\ulaeye}{\mathcal{U}}
\newcommand{\eyetop}{\mathrm{h}_+}
\newcommand{\eyebot}{\mathrm{h}_-}
\newcommand{\eyewidth}{\mathrm{w}}

\newcommand{\paramvar}{\rho}
\newcommand{\paramfreq}{\omega}
\newcommand{\paramspacing}{\Delta}
\newcommand{\spacingratio}{\delta}
\newcommand{\pvula}{\sigma}
\newcommand{\mfula}{\eta}

\newcommand{\paramdomain}{\mathcal P}
\newcommand{\antdomain}{\mathcal Z}
\newcommand{\sourcedomain}{\mathcal X}

\newcommand{\ndim}{d}
\newcommand{\paramdim}{q}

\newcommand{\foldindex}{p}

\newcommand{\antloc}{{\bs z}}
\newcommand{\sourceloc}{{\bs x}}
\newcommand{\testedloc}{{\mtest{\bs x}}}
\newcommand{\diffradius}{R}
\newcommand{\diffangle}{\theta}

\newcommand{\sourcek}{{k_\mathrm{c}}}
\newcommand{\sourcewavelen}{\lambda_\mathrm{c}}
\newcommand{\sourcecarrier}{f_\mathrm{c}}

\newcommand{\ucaradius}{R_\mathrm{uca}}
\newcommand{\aperture}{\psi}
\newcommand{\visualaperture}{\Omega}
\newcommand{\lenula}{L}

\newcommand{\xloc}{\mathrm{x}}
\newcommand{\yloc}{\mathrm{y}}
\newcommand{\xtest}{\tilde{\mathrm{x}}}
\newcommand{\ytest}{\tilde{\mathrm{y}}}
\newcommand{\dx}{d_{x}}
\newcommand{\prootula}{\bar{\paramvar}}
\newcommand{\prootulan}{\beta}
\newcommand{\prootulansec}{\beta^\mathrm{s}}

\newcommand{\xratio}{v}
\newcommand{\yratio}{u}
\newcommand{\xyratio}{w}

\newcommand{\seppar}{ ; \,}

%% file: heading/acronyms.tex
\newacronym{af}{AF}{ambiguity function}
\newacronym{abp}{BP}{beam pattern}
\newacronym{gbp}{GBP}{general beam pattern}
\newacronym{awgn}{AWGN}{additive white gaussian noise}
\newacronym{snr}{SNR}{signal-to-noise ratio}
\newacronym{mrt}{MRT}{maximum ratio transmission}
\newacronym{mrc}{MRC}{maximum ratio combining}

\newacronym{afr}{AFR}{aliasing-free region}
\newacronym{asod}{ASOD}{aliasing-safe operating domain}

\newacronym{ft}{FT}{Fourier transform}

\newacronym{aae}{AAE}{antenna array element}
\newacronym{ue}{UE}{user equipment}

\newacronym{nf}{NF}{near field}
\newacronym{ff}{FF}{far field}

\newacronym{ula}{ULA}{uniform linear array}
\newacronym{uca}{UCA}{uniform circular array}
\newacronym{mimo}{MIMO}{multiple-input multiple-output}
\newacronym{xla}{XL-array}{extremely large-scale array}

%% file: sections/1-Intro.tex
\section{Introduction}

With the recent advances towards 6G telecommunication networks and next-generation localization systems, the importance of \glspl{xla} has risen due to the increased spatial resolution and spectral efficiency they are anticipated to provide~\cite{li_near_field_2023, chen_6g_2024}.
This evolution is characterized by dimensions that fundamentally modify the operating regime of arrays. 
\rev{Consequently, the spherical nature of the wavefronts can no longer be ignored. 
This invalidates the conventional \gls{ff} condition---defined using the Fraunhofer distance---as it relies on a planar wavefront approximation.}
The \rev{\gls{nf}} regime in which \glspl{xla} operate enables a single array to transmit or receive steering signals that encode both range and angular information.
This feature supports advanced functionalities such as high-resolution positioning or beam pointing~\cite{demir_foundations_2021, wu_multiple_2023}.
Nonetheless, \rev{the \gls{nf} regime also gives rise to beampatterns with increased structural complexity, which complicates} the analytical characterization of system performance~\cite{sakhnini_near_field_2022, wang_extremely_2024, liu_near_field_2023}.

\glspl{xla} can be categorized based on their structural configurations, ranging from large co-located arrays (composed of numerous densely spaced antennas) to fully distributed \glspl{xla} that spread their constituent antennas over wide areas~\cite{lu_tutorial_2024}.
The latter is typically encountered in cell-free \gls{mimo} systems.
\rev{On the one hand, co-located \glspl{xla} commonly aim to meet the seminal half-wavelength antenna spacing criterion in order to avoid grating lobes. 
In the \gls{ff} regime, this phenomenon is well understood as the undesired illumination of multiple angles with similar power levels~\cite{lu_how_2021}.}
On the other hand, extending array dimensions while maintaining half-wavelength spacing introduces significant challenges in terms of computational complexity and physical constraints.
These limitations have motivated recent studies to investigate alternative structures (including sparse, modular, and distributed arrays) that relax this spacing requirement while maintaining a large array aperture~\cite{rodrigues_low_2020, zhou_sparse_2025, kosasih_spatial_dof_2025}.
Such designs require dealing with effects analogous to the grating lobes, but exhibiting significantly more intricate structure than their \gls{ff} counterparts.
The analysis of these effects constitutes the scope of this paper.

\medskip

In practice, this paper studies these effects as they appear in the \emph{\gls{af}}. The \gls{af} is a fundamental tool used to evaluate important indicators of localization performance, such as the resolution and the ambiguities~\cite{gui_generalized_2022, liu_near_field_2023}.
It is equivalent to the matched beampattern obtained with the \gls{mrt} beamformer~\cite{vandendorpe_positioning_2025}. 
The \gls{af} evaluates the correlations between all possible pairs of signals associated with locations observed by the array.  
\rev{In theory, the perfect \gls{af} would reduce to a Dirac delta function}, being non-zero only when matching twice the same location. 
\rev{In practice, however,} finite-sized arrays introduce non-orthogonality between distinct steering signals, typically resulting in \glspl{af} with oscillatory behaviors. 
The width of the main lobe characterizes the intrinsic resolution of the system. 
It indicates the minimum separation required between two sources to produce weakly correlated signals.
The secondary lobes generate estimation ambiguities.
It corresponds to a risk of confusion between distinct source locations, which, although far from one another, exhibit correlated signals.

Severe additional estimation ambiguities arise from the presence of grating lobes.
Mitigating their effect is essential to reach high performance in both single- and multi-source contexts.
In the \gls{ff}, it is often possible to obtain closed-form expressions of the \gls{af} with an angle-distance representation of the locations.
As a result, the \gls{ff} grating lobes are well-understood: antenna spacing exceeding half the wavelength introduces angular repetitions. 
Thanks to this understanding, \rev{their presence can be mitigated by restricting the source} locations to an angular sector as seen by the array.
Alternatively, several other techniques have been suggested in the literature to address grating lobe mitigation in the \gls{ff}.
In~\cite{zhuang_coherent_2008}, a sparse \gls{mimo} system composed of several linear subarrays transmitting at distinct frequencies is considered. 
The authors in~\cite{krivosheev_grating_2010} reduce the grating lobes by leveraging variations in subarray spacing, orientation, and density.
In \cite{gao_integrated_2023}, the authors resolve angular repetitions using a compressed sensing approach, constrained to an angular sector.

To mitigate the effect of grating lobes in the \gls{nf} regime, it is essential to mathematically characterize their geometric structure in the \gls{af}.
\rev{This task is particularly challenging because the \gls{nf} regime introduces nonlinear spatial phase variations in steering signals, which generally preclude the derivation of exact closed-form expressions of the \gls{af}.}
Recent approaches have largely relied on angle-distance domain analysis, leveraging \emph{simplifying assumptions} to obtain approximate closed-form expressions of the \gls{af} for specific array topologies. 
For example, the authors in \cite{li_near_field_2023} exploit the unique regular structure of modular \glspl{xla} and define a sub-array-based uniform spherical wave model that assumes an inner \gls{ff} regime within each module. 
In \cite{kosasih_spatial_dof_2025}, the achievable beamwidth that maintains a low grating lobe level is investigated in a similar modular \gls{xla} context. 
Additionally, \cite{zhou_sparse_2025} exploits Fresnel approximations to evaluate the beampatterns for linear sparse arrays and extended coprime arrays.

While these approaches provide valuable insights, they rely on context-specific simplifications that limit their generality. 
Moreover, they exploit definitions of \gls{nf} grating lobes that heuristically \rev{extend those used in the \gls{ff} regime}, thereby inherently lacking a formal theoretical foundation in the \gls{nf} regime.
\rev{In the \gls{ff} regime, grating lobes admits} three \emph{equivalent} definitions:
\begin{enumerate}
    \item Secondary lobes with the same amplitude as the main lobe.
    \item The effect of pairs of distinct locations (or angles) that produce identical steering signals up to a constant phase shift.
    \item The effect of locations yielding discrete steering signals affected by spatial aliasing.
\end{enumerate}
In the \gls{nf}, the first two definitions rarely (and presumably never) occur because the non-linear phase structure prevents exact repetitions of steering signals.
This behavior is observed in~\cite{li_near_field_2023, lu_tutorial_2024, zhang_near_field_2025} and in the \glspl{af} numerically computed in our paper.
\rev{By contrast}, defining the grating lobes as spatial \emph{aliasing artifacts} remains formally valid in the \gls{nf} regime.
\rev{Building upon this definition, we develop} a framework that is both theoretically grounded in the structural origins of grating lobes and directly consistent with established \gls{ff} conventions.

Our novel framework provides the key advantage of determining grating lobe properties without requiring the challenging derivation of closed-form expressions for the \gls{af}.
Exploiting a \emph{spatial frequency-domain} representation of signals, \rev{we can identify the set of locations} that generates the spectral folding associated with visible grating lobes in the \gls{af}.
Furthermore, the abstract geometric formulation of our methodology enables its applicability to a wide range of array topologies.

As a result, our framework supports the systematic design of \gls{nf} \gls{xla} systems that operate robustly with respect to grating lobes.
Indeed, once the structure of aliasing effects is understood, antenna positions and spacings can be related to regions of aliasing-free operations.
Building on this principle, the present paper departs from prior studies and proposes a theoretical construction of aliasing-safe operational areas for \glspl{xla}. 

\medskip

To \rev{establish} this new framework, we compressively capture the spatial spectral content of steering signals by exploiting the concept of \emph{local} spatial frequencies. 
\rev{This concept has already been formalized} in multiple studies conducted in the context of channel degrees of freedom~\cite{ding_degrees_2022, ding_spatial_2024, kosasih_spatial_dof_2025}.
Similarly, \gls{nf} steering signals can be interpreted as \emph{spatial chirp}.
This concept, notably reported in~\cite{swindlehurst_passive_1988, qiu_doa_2018, jian_fractional_2024}, aligns with the standard instantaneous frequency representation of chirp signals in the time domain.

\medskip

This paper therefore extends the preliminary findings of \cite{monnoyer_chirp_2025} and proposes a systematic methodology to characterize the aliasing structure that arises in the \gls{af} of \glspl{xla} operating in the \gls{nf} regime.
It also derives practical design guidelines for \glspl{xla}.
Using a local \rev{spatial-frequency} (or ``chirp-based") representation of steering signals, the proposed framework is connected to established results in the \gls{ff} regime.

Consequently, the contributions of this paper are:
\begin{itemize}
    \item A mathematical formalism to define the \gls{nf} grating lobes as the effect of ``spatial aliasing artifacts"; 
    \item A novel framework enabling the systematic mathematical description of the geometric structures induced by these artifacts on the \gls{af};
    \item Practical guidelines for designing arrays that operate within aliasing-safe regions;
    \item A compressive modeling of the steering signal spectrum that connects the \gls{nf} analysis with \gls{ff} results;
    \item Interpretable closed-form expressions describing the complex grating lobe structures for canonical \glspl{ula} and \glspl{uca}.
\end{itemize}

The remainder of this paper is organized as follows.
Section \ref{sec:model} introduces the theoretical model considered in this paper. 
Section \ref{sec:framework} formalizes the concepts of the \gls{af} and aliasing artifacts, and presents the proposed framework to understand their structure.
Next, section \ref{sec:chirp} introduces a spectral modelling of the steering signals that connects the proposed framework to established \gls{ff} principles.
This section also details the practical guidelines that originate from our framework.
Finally, sections \ref{sec:ula} and \ref{sec:uca} apply the proposed theoretical framework to specific \gls{ula} and \gls{uca} topologies, providing closed-form expressions that explain the structure of the resulting grating lobes.


\begin{figure*}[!t]
    \centering
    \includegraphics[width=0.47\linewidth]{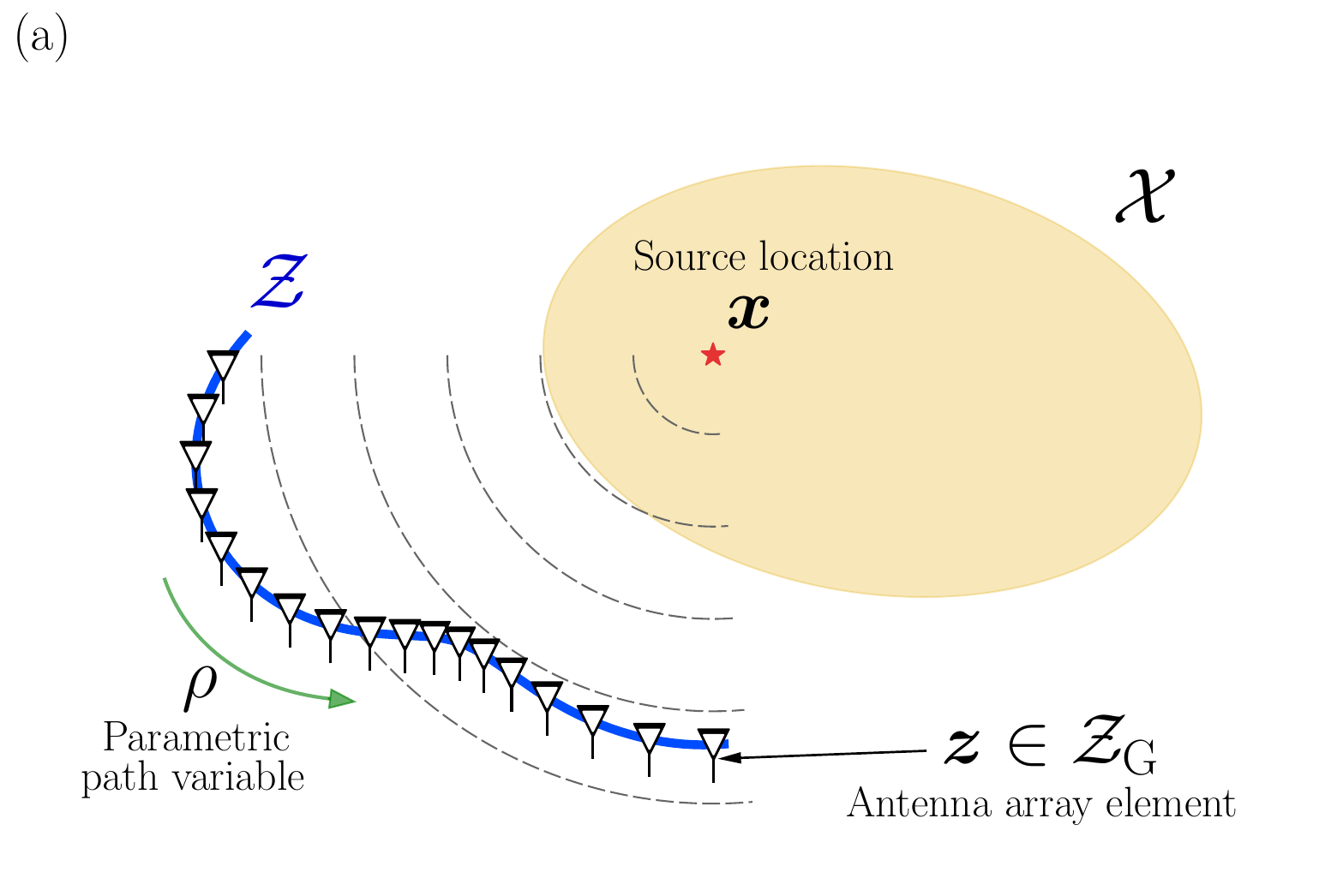}
    \hspace{0.01\linewidth}
    \includegraphics[width=0.47\linewidth]{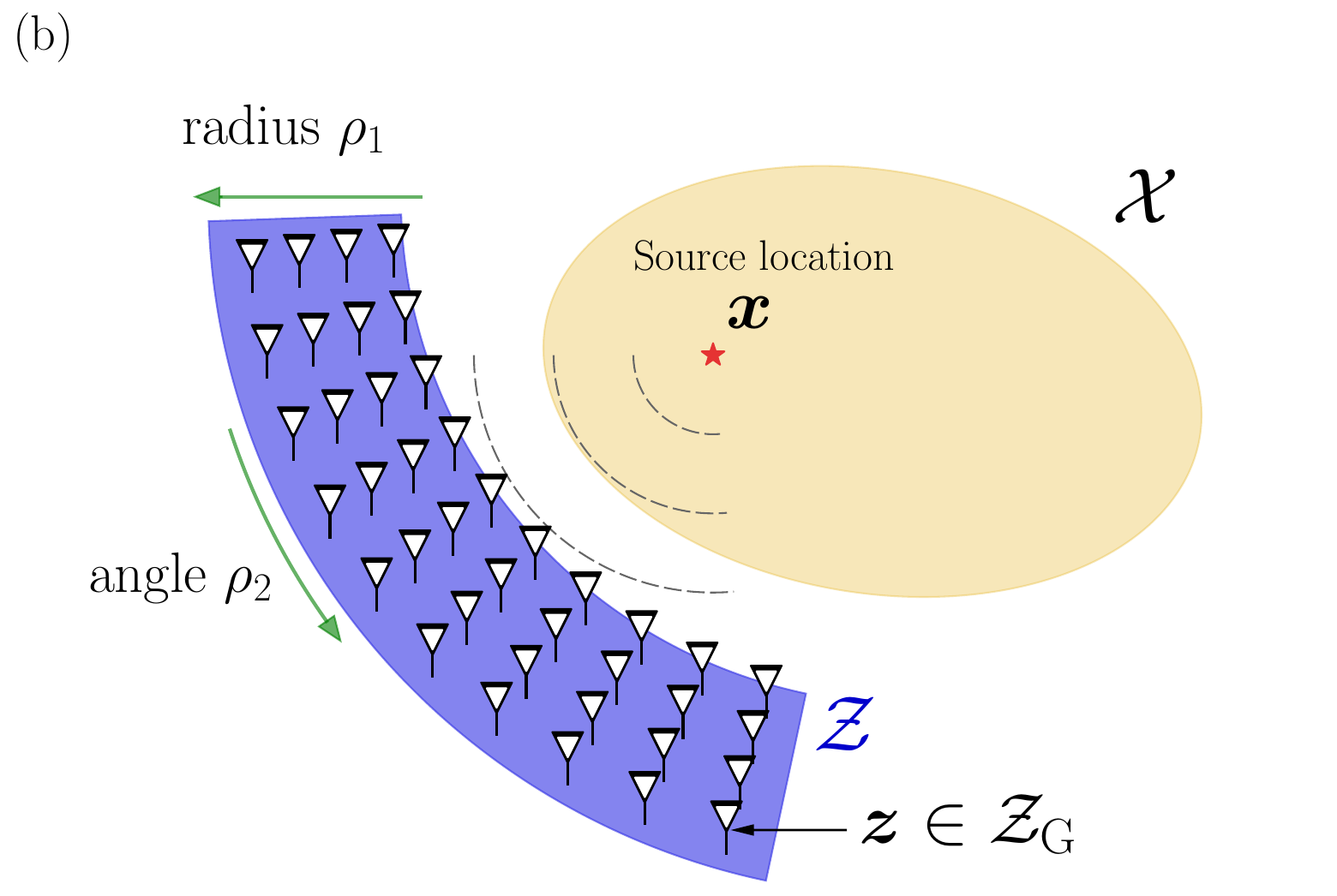}
    \caption{Schematic representation of the uplink scenario studied in this paper. 
    (a) A case where the antenna array is defined over a curve $\antloc$ parameterized by a scalar path variable $\paramvar$.
    (b) A case where the antenna array is parameterized with two path variables. Representation of the simple case of the polar description.}
    \label{fig:scenario-standard}
\end{figure*}

\subsection*{Mathematical Notations}
In the current paper, the imaginary unit is $j = \sqrt{-1}$ and $\cel$ is the speed of light.
The operator $\otimes$ denotes the convolution product.
The symbol $\seppar$ separating inputs in functions, as $f(x \seppar y)$, indicates its functional dependence on the variable $x$ and parameterization by $y$.
Bold lowercase letters denote vectors, and caligraphic-style uppercase letters, like $\mathcal P$, are sets.
The notations and $\|\bs a\|$ and $\angle \bs a$ respectively are the $\ell_2$-norm and the angle of the vector $\bs a$.
Given the vector $\bs a$ and the scalar factor $b$, the set $b \cdot \mathcal P + \bs a$ is identical to $\mathcal P$ but rescaled by $b$ and translated by $\bs a$.

%% file: sections/2-Model.tex
\subsection{System Model} 
\label{sec:model}

The content of this paper is formulated for a canonical uplink scenario in which a static \gls{ue}, located in $\sourceloc$, transmits a unit-energy baseband signal $\sigtx(t)$ modulated \rev{onto a carrier with} frequency $\sourcecarrier$. 
As illustrated in Figure~\ref{fig:scenario-standard}, the transmitted signal is \rev{received} by an \gls{xla}.
The antennas are located on the grid $\antgrid$ that discretizes the continuous-space domain $\antdomain$.
Note that our system model allows $\antgrid$ to represent \emph{any} discrete subset of $\antdomain$, and is therefore not restricted to uniform planar grids.
\rev{The receiver performs network-based positioning, assuming that the source location $\sourceloc$ belongs to an \emph{operating domain} $\sourcedomain \subseteq \bb R^\ndim$.} 

Following physical-optics principles, the \gls{aae} located in $\antloc\in\antdomain$ measures, up to an \gls{awgn}, the baseband equivalent signal
\begin{equation}
    \label{eq:receiver-signal}
    \sigrx(t, \antloc \seppar \sourceloc) = 
    \frac{e^{-j \sourcek \|\antloc - \sourceloc\|}}{\|\antloc - \sourceloc\|} 
    \sigtx\big(t - \tfrac{\|\antloc - \sourceloc\|}{\cel}\big),
\end{equation}
where $\sourcek = 2\pi \sourcecarrier/\cel = 2\pi/\sourcewavelen$.


To simplify mathematical developments, we use a normalized formulation with respect to the total received energy
\begin{equation}
    \energy(\sourceloc) 
        =
        \int_{\antdomain}
        \|\antloc - \sourceloc\|^{-2} 
        \: \mathrm d\antloc.
\end{equation}
Using the above, we can now write
\begin{equation}
    \sigrx(t, \antloc \seppar \sourceloc) = 
    \energy^{\frac{1}{2}}(\sourceloc)\,\sig(\antloc \seppar \sourceloc)\, 
    \sigtx\big(t - \tfrac{\|\antloc - \sourceloc\|}{\cel}\big),
\end{equation}
where the wave propagation coefficients, or the \emph{steering signal},
\begin{equation}
    \label{eq:steering-signal}
    \sig(\antloc \seppar \sourceloc) 
    =
    \amplitudefun(\antloc\seppar \sourceloc) \exp\big(-j \phasefun(\antloc\seppar \sourceloc)\big),
\end{equation}
represents the energy-normalized \rev{attenuation and phase} received at the antenna located in position $\antloc$, parameterized by the given source location $\sourceloc$, with
\begin{align}
    \label{eq:phasefun}
    \phasefun(\antloc \seppar \sourceloc) &= \sourcek \|\antloc - \sourceloc\|,
    \\
    \label{eq:amplitudefun}
    \amplitudefun(\antloc \seppar \sourceloc) &= \big(\energy^{\frac{1}{2}}(\sourceloc) \|\antloc - \sourceloc\|\big)^{-1}.
\end{align}
By definition, the steering signal $\sig(\antloc \seppar \sourceloc)$ \rev{has} unit energy regardless of the value of $\sourceloc$.

The current paper assumes a narrowband signal $\sigtx(t)$, \rev{therefore focusing exclusively on} the wave propagation coefficients described by $\sig(\antloc \seppar \sourceloc)$.
\rev{Consideration of wideband aspects and extensions to moving \glspl{ue} is left for future work.}


In the \gls{nf} regime, the phase $\phasefun(\antloc \seppar \sourceloc)$ \rev{varies \emph{non-linearly} across $\antloc$}, enabling beam pointing at a specific location. 
While the \emph{continuous-space} steering signal $\sig(\antloc \seppar \sourceloc)$ \rev{achieves this} focusing, its \emph{discrete-space} counterpart (obtained from the finite set of antennas in $\antgrid$) may introduce grating lobes, resulting in localization ambiguities.
As conventional \gls{ff} interpretations of the grating lobes ignore the non-linear phase structure, they do not apply in the \gls{nf}. This is a gap our framework addresses in the following sections.

\subsection*{Parametric representation}

\rev{This paper adopts a generalized parametric model based on a \emph{bijective function} $\paramtf$ mapping a compact parametric domain $\paramdomain \in \bb R^\paramdim$ onto the antenna domain $\antdomain$, such that}
\begin{equation}
    \antdomain = \{ \paramtf(\paramvar) : \paramvar \in \paramdomain \}.
\end{equation}
For example, Figure \ref{fig:scenario-standard}a shows a 1D curved domain $\antdomain$ parameterized by the path variable $\paramvar$. Figure \ref{fig:scenario-standard}b shows a case where $\paramvar$ is a polar description of $\antloc$.
This parameterization enables to
\begin{enumerate}[label={(\roman*)}]
    \item represent a wide range of array topologies within a single framework, including canonical \glspl{ula}, \glspl{uca}, and paving the way for more complex structures,
    \item assume only uniform Cartesian sampling of the parametric domain $\paramdomain$, denoted by $\paramgrid$, without loss of generality in the definition of $\antgrid$.
\end{enumerate}
Any grid $\antgrid$ can be described by a uniform Cartesian grid $\paramgrid$ via a bijection $\paramtf$, preserving the generality of our approach.
This mapping allows grating lobes originating from any set of locations $\antgrid$ to be analyzed using standard Fourier theory on $\paramgrid$.

\rev{For clarity}, \rev{this paper expresses its results} only for $\paramdim=1$, 
as represented in Figure~\ref{fig:scenario-standard}a, with $\paramspacing$ denoting the sampling step generating $\paramgrid$ from $\paramdomain$.
The extension to $\paramdim>1$ can be obtained by following the steps of this paper with multi-dimensional Fourier transforms, in which the spectral folding must be considered separately in each dimension.

By substituting $\antloc = \paramtf(\paramvar)$ in the steering signal model, the analysis focuses on the signal  
\(
    \sig\big(
        \paramtf(\paramvar) \seppar \sourceloc
    \big)
\)
as a function of $\paramvar$.

%% file: sections/3-Aliasing.tex
\rev{
\section{Aliasing artifacts in the ambiguity function}
\label{sec:framework}

In this section, we define the \gls{af} for the \gls{nf} sensing scenario introduced in Section \ref{sec:model}. 
We establish a formalism to characterize the \gls{nf} grating lobes as the result of an aliasing process. 
\rev{To illustrate the methodology, we present a toy example consisting of a horizontal \gls{ula} of length $1000\sourcewavelen$ centered at the origin.}
Defining the trivial transformation $\paramtf_\mtoy(\paramvar) := [\paramvar, 0]^\top$, we build the grid $\antgrid$ by taking values of $\paramvar$ in the uniform grid $\paramgrid$ with a spacing $\paramspacing = 10\sourcewavelen$.
For example, with a high carrier frequency $\sourcecarrier = 60$GHz, this corresponds to a co-located 5m-long linear array with a $5$cm antenna spacing.
Alternatively, at a lower carrier frequency $\sourcecarrier = 200$MHz, it can also correspond to a \emph{distributed} array composed of 100 access points spaced 15m apart.
We fix the user location in $\sourceloc = [0, 600\sourcewavelen]$, centered in front of the array, in its \gls{nf} operational region.

\subsection{Near-field ambiguity functions}

The \gls{af} is an important tool \rev{for evaluating} the structural properties of a sensing system. 
Motivated by a maximum likelihood estimation approach in the presence of \gls{awgn}, it is defined as the noise-free matched-filter response of the sensing system for a \emph{candidate} position $\testedloc$, while the \emph{true} source's position is $\sourceloc$.
For a \emph{wideband} localization system, its standard expression is
\begin{equation}
    \label{eq:def-spatial-ambifun-band}
    \ambifunband(\testedloc, \sourceloc)
    :=
    \sum_{\antloc \in \antgrid} 
    \int_{-\infty}^{\infty}
    \bar\sigrx^*(t, \antloc \seppar \testedloc)
    \:
    \bar\sigrx(t, \antloc \seppar \sourceloc)
    \,\mathrm{d} t,
\end{equation}
where $\bar\sigrx(t, \antloc \seppar \sourceloc) = \energy^{-\frac{1}{2}}(\sourceloc)\sigrx(t, \antloc \seppar \sourceloc)$. 

In practice, the \gls{af} is inherently tied to the probability of erroneously detecting $\testedloc$ instead of the true location $\sourceloc$. 
When multiple sources must be located, it also \rev{quantifies} the mutual interference between signals received from the different \rev{\glspl{ue}}. 
\rev{Higher values of $\ambifunband(\testedloc, \sourceloc)$ indicate a higher probability of failing to resolve the two locations.}
This situation typically arises, by definition, when $\testedloc$ and $\sourceloc$ are separated by less than the array’s intrinsic resolution. 
However, the grating lobes reveal pairs of \emph{further apart} locations that can also be unresolvable.
This emphasizes the need to study their properties.

\begin{figure}[t]
    \centering
    \includegraphics[width=\linewidth]{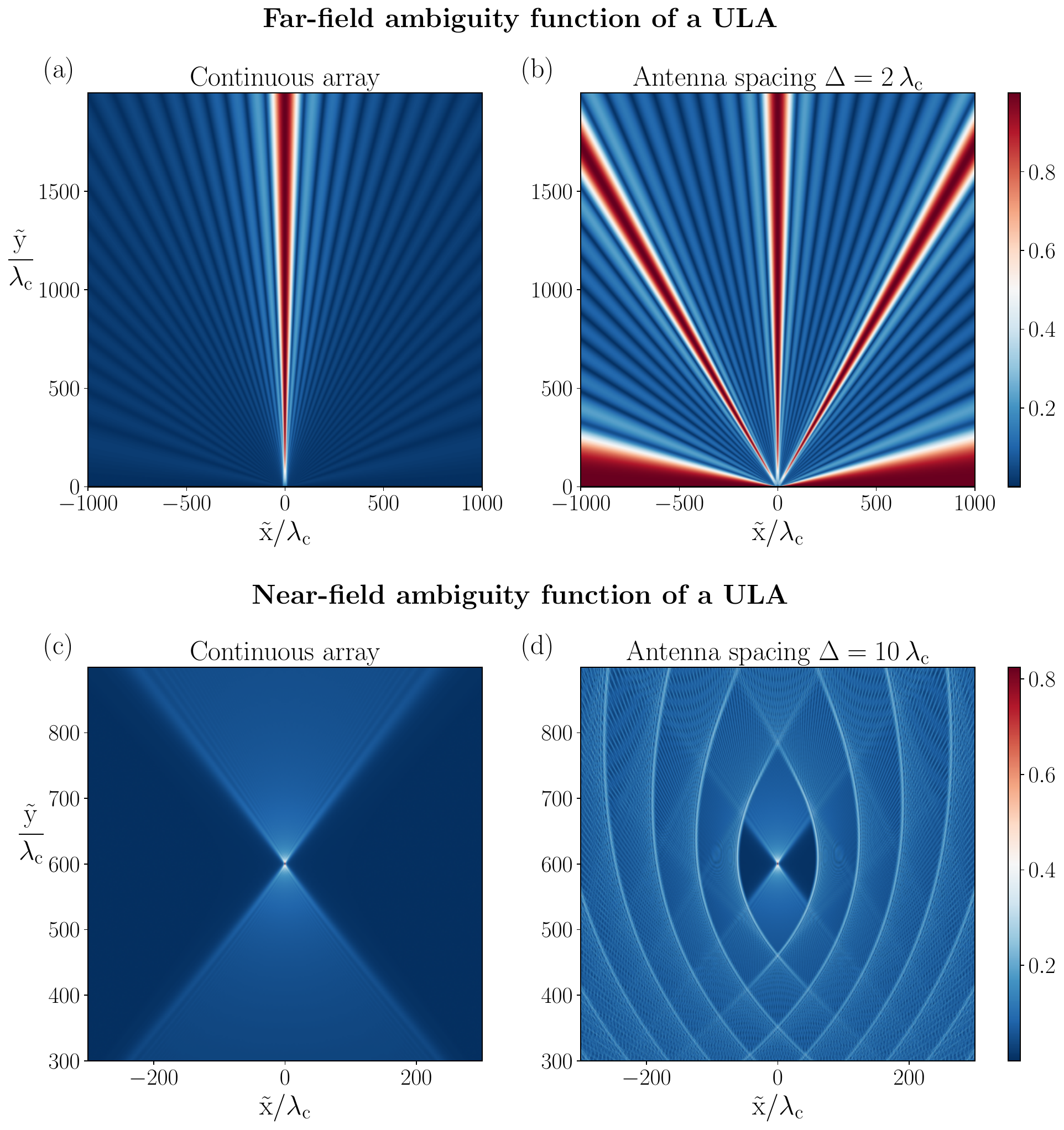}
    \caption{Ambiguity functions for (top) a horizontal \gls{ula} with length 20$\sourcewavelen$ operating in the \gls{ff}, and (bottom) the \gls{ula} from our toy example in the \gls{nf}, as a function of $\testedloc = [\xtest, \ytest]^\top$ given the fixed location $\sourceloc = [\xloc, \yloc] = [0, 600]\sourcewavelen$. 
    It shows in (a)-(c) the continuous-space $|\ambifun(\testedloc, \sourceloc)|$ and in (b)-(d) the discrete-space $|\sambifun(\testedloc, \sourceloc)|$, with $\paramspacing$ chosen for improved visibility of the grating lobes.}
    \label{fig:AF-toy}
\end{figure}

\medskip

In the narrowband settings of this paper, the \gls{af} reduces to the correlation function between possible pairs of steering signals, 
providing the theoretical signature of the antenna array.
\rev{For the continuous steering signal, it is expressed as} 
\begin{equation}
    \label{eq:cs-ambifun}
    \ambifun(\testedloc, \sourceloc) 
    :=
    \int_{\paramdomain}
    \matchfun(\paramvar \seppar \testedloc, \sourceloc)
    \:\mathrm{d} \paramvar,
\end{equation}
where we define the \emph{matched signal} as
\begin{equation}
    \label{eq:matching-fun}
    \matchfun(\paramvar \seppar \testedloc, \sourceloc) := 
    \sig^*\big(\paramtf(\paramvar) \seppar \testedloc\big)
    \,
    \sig\big(\paramtf(\paramvar) \seppar \sourceloc\big),
\end{equation}
for the simplicity of the notations throughout this section. 
Considering the discrete-space steering signal observed by the antennas in $\antgrid$, the discrete-space \gls{af} is expressed as
\begin{align}
    \label{eq:ds-ambifun}
    \sambifun(\testedloc, \sourceloc) &:= 
    \paramspacing
    \sum_{\paramvar \in \paramgrid} 
    \matchfun(\paramvar \seppar \testedloc, \sourceloc).
\end{align}

The \gls{af} behaves differently in the \gls{ff} and \gls{nf} regimes.
Figure~\ref{fig:AF-toy} illustrates this contrast by showing continuous- and discrete-space \glspl{af} for \glspl{ula} operating in both contexts.
Compared to its \gls{ff} counterpart, the \gls{nf} \gls{af} \rev{achieves a high resolution in both angle and range.}

With the large antenna spacing used in Figures~\ref{fig:AF-toy}b and \ref{fig:AF-toy}d, the discrete-space \glspl{af} display grating lobes.
Their well-understood angular structure in the \gls{ff} regime appears in Figure~\ref{fig:AF-toy}b, whereas Figure~\ref{fig:AF-toy}d \rev{shows the more complex geometry observed in the \gls{nf} regime.}
By developing on the \emph{aliasing} nature of the \gls{nf} grating lobes, this paper mathematically characterizes their origin and their intricate geometrical structure.
Although the resolutions of the main lobe and the grating lobes are additional important factors affecting both localization and communication performance, extending our framework to include this aspect is left for future studies.

\medskip

We finally highlight the mathematical connection between the narrowband \gls{af} and the concept of beampattern, which is commonly used to study the performance of communication systems. 
The beampattern is classically defined as the set of correlations between each steering vector and a \emph{generic} beamformer, \ie one not restricted to the set of steering vectors.
As discussed in~\cite{vandendorpe_positioning_2025}, the \gls{af} in \eqref{eq:ds-ambifun} corresponds precisely to the beampattern obtained with beamforming weights tailored for the specific \gls{mrt} (or \gls{mrc}) strategy.

\subsection{Aliasing artifacts}
\label{sec:aliasing}

We can now develop the spectral folding process that generates ``aliasing artifacts" in $\sambifun(\testedloc, \sourceloc)$, thereby defining the \gls{nf} grating lobes.
To this end, \rev{we exploit the mathematical connection between \eqref{eq:cs-ambifun} and \eqref{eq:ds-ambifun}.
The continuous-space \gls{af} in \eqref{eq:cs-ambifun} can be interpreted as evaluating the \gls{ft}}, $\matchspectrum(\paramfreq \seppar \mtest\sourceloc, \sourceloc)$, of the matched signal $\matchfun(\paramvar \seppar \testedloc, \sourceloc)$ for $\paramfreq = 0$.
Indeed, given that
\begin{equation}
    \label{eq:matched-spectrum}
    \matchspectrum(\paramfreq \seppar \mtest\sourceloc, \sourceloc)
    =
    \int_{\paramdomain} 
    \matchfun(\paramvar \seppar \testedloc, \sourceloc) 
    e^{-j\paramfreq \paramvar}
    \:\mathrm d \paramvar,
\end{equation}
we have
\begin{align}
    \label{eq:cs-ambifun-equality}
    \ambifun(\testedloc, \sourceloc) &= \matchspectrum(0 \seppar \testedloc, \sourceloc).
\end{align}
\rev{Figure~\ref{fig:toy-spectrum}a illustrates the spectra $\matchspectrum(\paramfreq \seppar \mtest\sourceloc, \sourceloc)$ for several tested locations $\testedloc$.}
The reference spectrum obtained when $\testedloc = \sourceloc$ exhibits a \rev{unit-height peak at $\paramfreq=0$}, yielding $\ambifun(\sourceloc, \sourceloc) = 1$.
As $\testedloc$ becomes further away from $\sourceloc$, the spectral energy in $\matchspectrum(\paramfreq \seppar \mtest\sourceloc, \sourceloc)$ is both frequency-spread and shifted, lowering its value in $\paramfreq=0$ and thus decreasing $\ambifun(\testedloc, \sourceloc)$.

\rev{The discrete-space \gls{af}, given in \eqref{eq:ds-ambifun}, analogously amounts to the \gls{ft} of the discretized matched signal, in $\paramfreq = 0$.}
Uniformly sampling $\matchfun$ causes repetitions in its spectrum, which directly lead to 
\begin{align}
    \label{eq:ds-ambifun-equality}
    \sambifun(\testedloc, \sourceloc) &= \sum_{\foldindex \in \bb Z} \matchspectrum( \tfrac{2\pi}{\paramspacing} \foldindex \seppar \testedloc, \sourceloc) \\
    \label{eq:ds-ambifun-equality-2}
    &= \ambifun(\testedloc, \sourceloc) + \sum_{\foldindex \in \bb Z \backslash \{0\}} \matchspectrum( \tfrac{2\pi}{\paramspacing} \foldindex \seppar \testedloc, \sourceloc).
\end{align}
The right-side summation of \eqref{eq:ds-ambifun-equality-2} isolates the aliasing artifacts affecting $\sambifun(\testedloc, \sourceloc)$ relatively to the continuous-space reference $\ambifun(\testedloc, \sourceloc)$.
These aliasing terms are strictly zero when the spectrum $\matchspectrum(\paramfreq \seppar \mtest\sourceloc, \sourceloc)$ spans frequencies that are lower, in absolute value, than the sampling frequency $2\pi/\paramspacing$.
This observation is similar to the Shannon-Nyquist sampling theorem, but is relaxed by a factor of 2.
This factor stems from the specificity that aliasing artifacts appear in $\sambifun(\testedloc, \sourceloc)$ only when spectral folding affects $\matchspectrum$ in $\paramfreq = 0$.
\rev{Folding at other frequencies has no consequence on the discrete-space \gls{af}.}

Formally, defining the maximum frequency content, coined the ``strict band limit", of the matched signal $\matchfun(\paramvar\seppar \testedloc, \sourceloc)$, as 
\begin{equation}
    \label{eq:def-band-limit}
    \bandlimstrict(\testedloc, \sourceloc) = 
    \max\{|\paramfreq|: 
    |\matchspectrum(\paramfreq \seppar \mtest\sourceloc, \sourceloc)| > 0\},
\end{equation}
this \emph{sufficient} condition is given in Lemma~\ref{lem:strict-no-aliasing}.

\begin{lemma}
    \label{lem:strict-no-aliasing}
    Given a pair of locations $\testedloc$ and $\sourceloc$, if the band limit satisfies
    \begin{equation}
        \label{eq:folding-condition}
        \bandlimstrict(\testedloc, \sourceloc) \leq 2\pi \paramspacing^{-1}, 
    \end{equation}
    then the equality $\sambifun(\testedloc, \sourceloc) = \ambifun(\testedloc, \sourceloc)$ holds.
\end{lemma}
\begin{proof}
    By construction of \eqref{eq:def-band-limit}, the condition \eqref{eq:folding-condition} directly nullifies all the terms in the right-side summation of \eqref{eq:ds-ambifun-equality-2}.
\end{proof}
Reciprocally, for a given location $\sourceloc$, aliasing artifacts can appear in the discrete-space \gls{af} for the tested locations $\testedloc$ that do \emph{not} satisfy \eqref{eq:folding-condition}.
As a result, Lemma~\ref{lem:strict-no-aliasing} is the key to determining the set of $\testedloc, \sourceloc$ that are unaffected by grating lobes in $\sambifun(\testedloc, \sourceloc)$.

\begin{figure}[t]
    \centering
    \includegraphics[width=\linewidth]{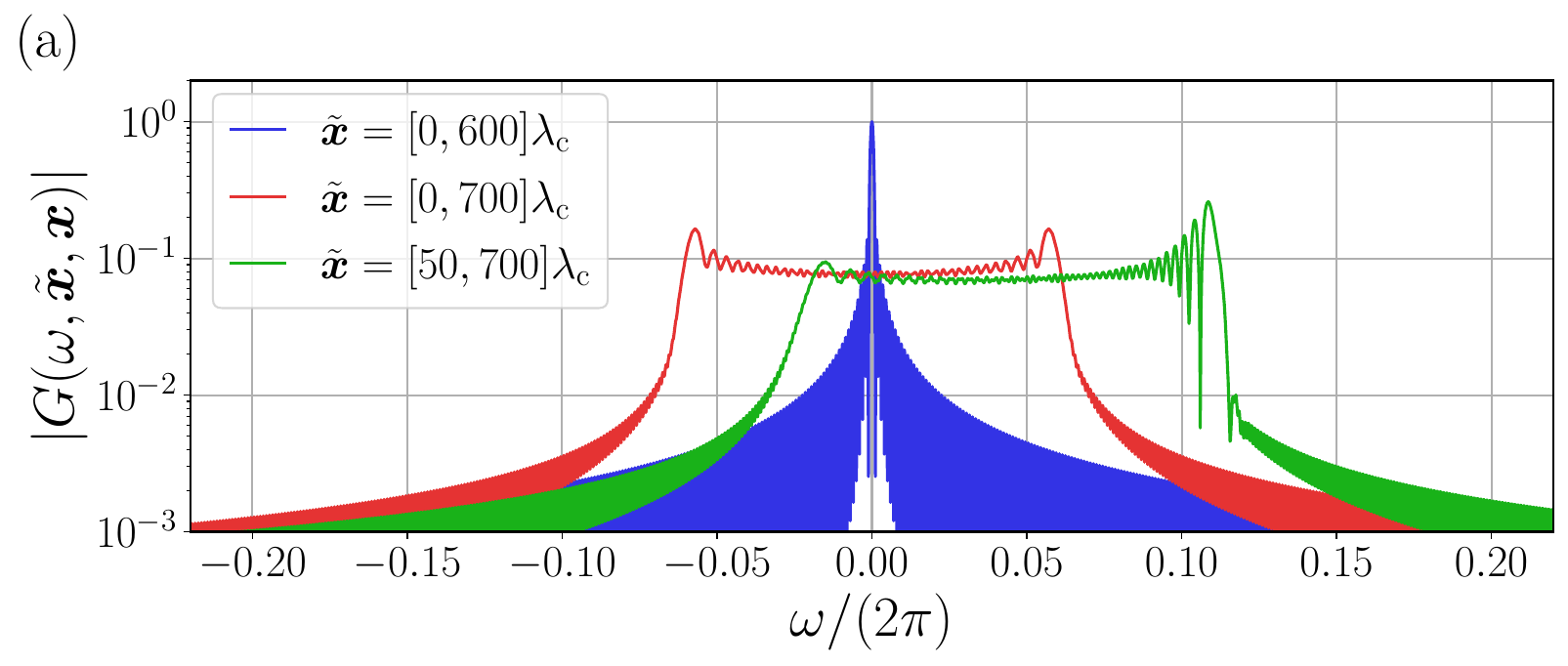}
    \includegraphics[width=\linewidth]{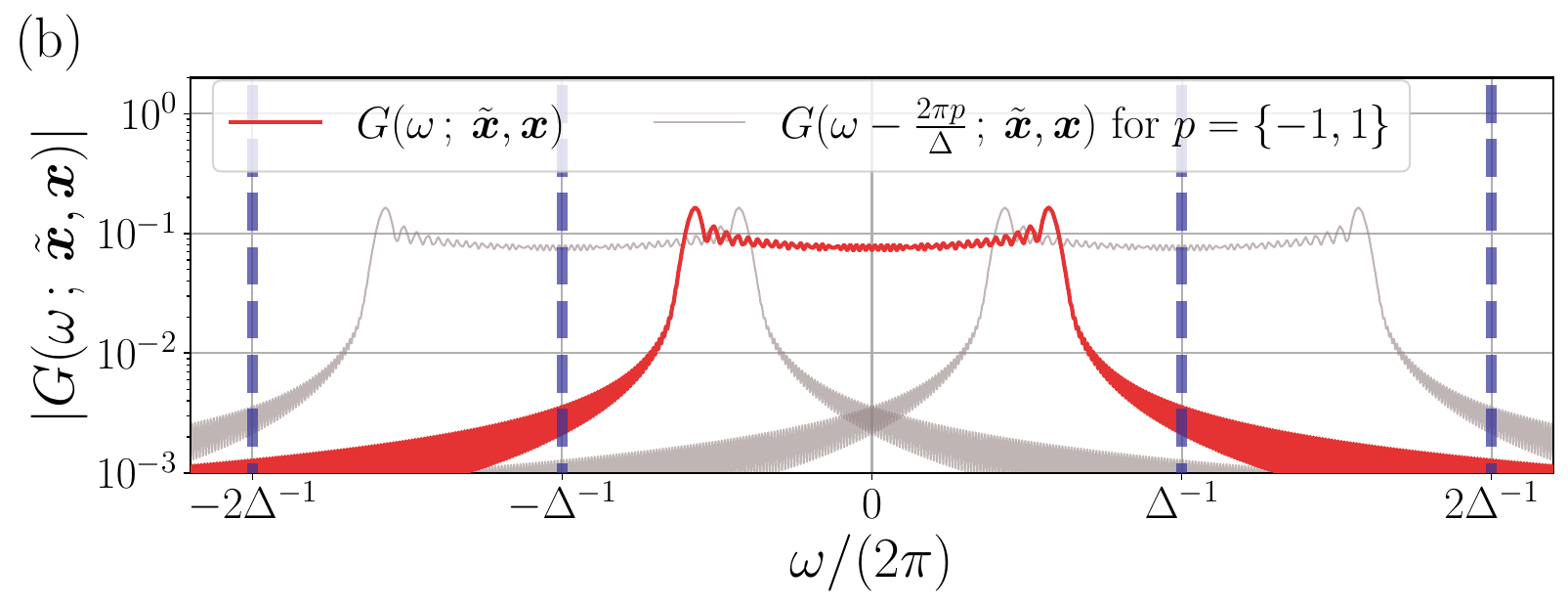}
    \caption{Matched spectrum's amplitude $|\matchspectrum(\paramfreq \seppar \mtest\sourceloc, \sourceloc)|$ corresponding to the toy example for multiple tested locations $\testedloc$, given $\sourceloc = [0, 600]\sourcewavelen$.}
    \label{fig:toy-spectrum}
\end{figure}

In practice, however, the strict band limit used in Lemma~\ref{lem:strict-no-aliasing} is typically infinite, as illustrated in Figure~\ref{fig:toy-spectrum}.
Instead, a common practice is to define} the band limit in a \emph{soft} manner, using a small tolerance threshold $\threshband$ rather than a strict zero in \eqref{eq:def-band-limit}. 
Yet, this intuitive approach lacks consistency with established \gls{ff} conventions, and is sensitive to the choice of $\threshband$.
As detailed in Section~\ref{sec:chirp}, we address this matter by introducing a chirp-based compressive representation of the spectral content of $\matchspectrum(\paramfreq \seppar \mtest\sourceloc, \sourceloc)$.
This representation bridges our \gls{nf} framework with the classical \gls{ff} interpretation of grating lobes, and supports design guidelines for \glspl{xla}.
\medskip

Figure~\ref{fig:toy-spectrum}b illustrates the \rev{spectral repetitions for} $\testedloc = [0, 700]\sourcewavelen$, at every multiple of $\frac{2\pi}{\paramspacing}$.
Although \rev{significant folding occurs in some part} of the spectrum, the zero-frequency is barely affected, resulting in no visible artifacts in the discrete-space \gls{af} in $\testedloc = [0, 700]\sourcewavelen$ (see Figure~\ref{fig:AF-toy}b).
\rev{Intuitively, Lemma~\ref{lem:strict-no-aliasing} states that if the spectrum's ``significant" support remains entirely within the $\frac{2\pi}{\paramspacing}$ limits, then $\sambifun(\testedloc, \sourceloc)$ remains aliasing-free.}
\rev{Importantly}, outside the specific \gls{ff} case, the \gls{af} itself is \emph{not} directly repeated by the sampling of $\matchfun$. 
Rather, the geometry of the artifacts altering $\sambifun(\testedloc, \sourceloc)$ is dictated, in each direction around $\sourceloc$, by the spread and shift behaviors of $\matchspectrum$ toward the $\tfrac{2\pi}{\paramspacing}$ boundaries as $\testedloc$ moves away from $\sourceloc$.

\begin{figure}
    \label{fig:toy-spectrum-spread}
    \centering
    \includegraphics[width=\linewidth]{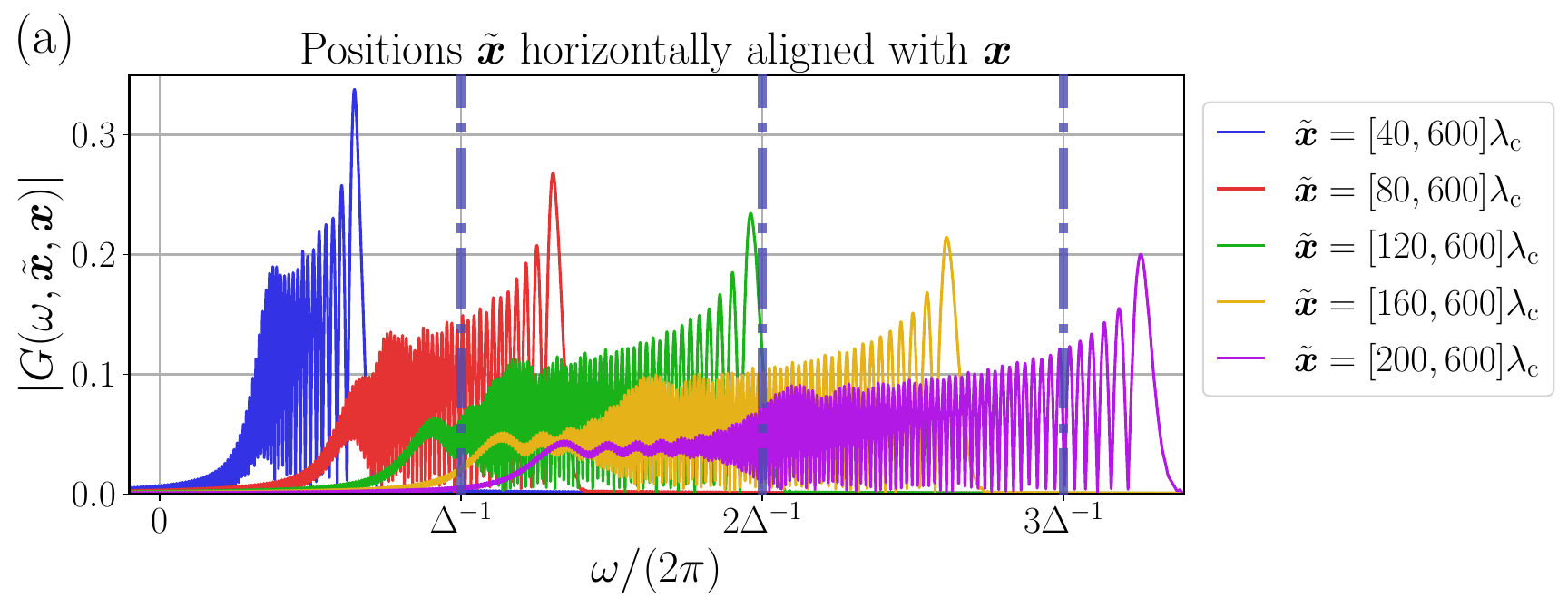}
    \includegraphics[width=\linewidth]{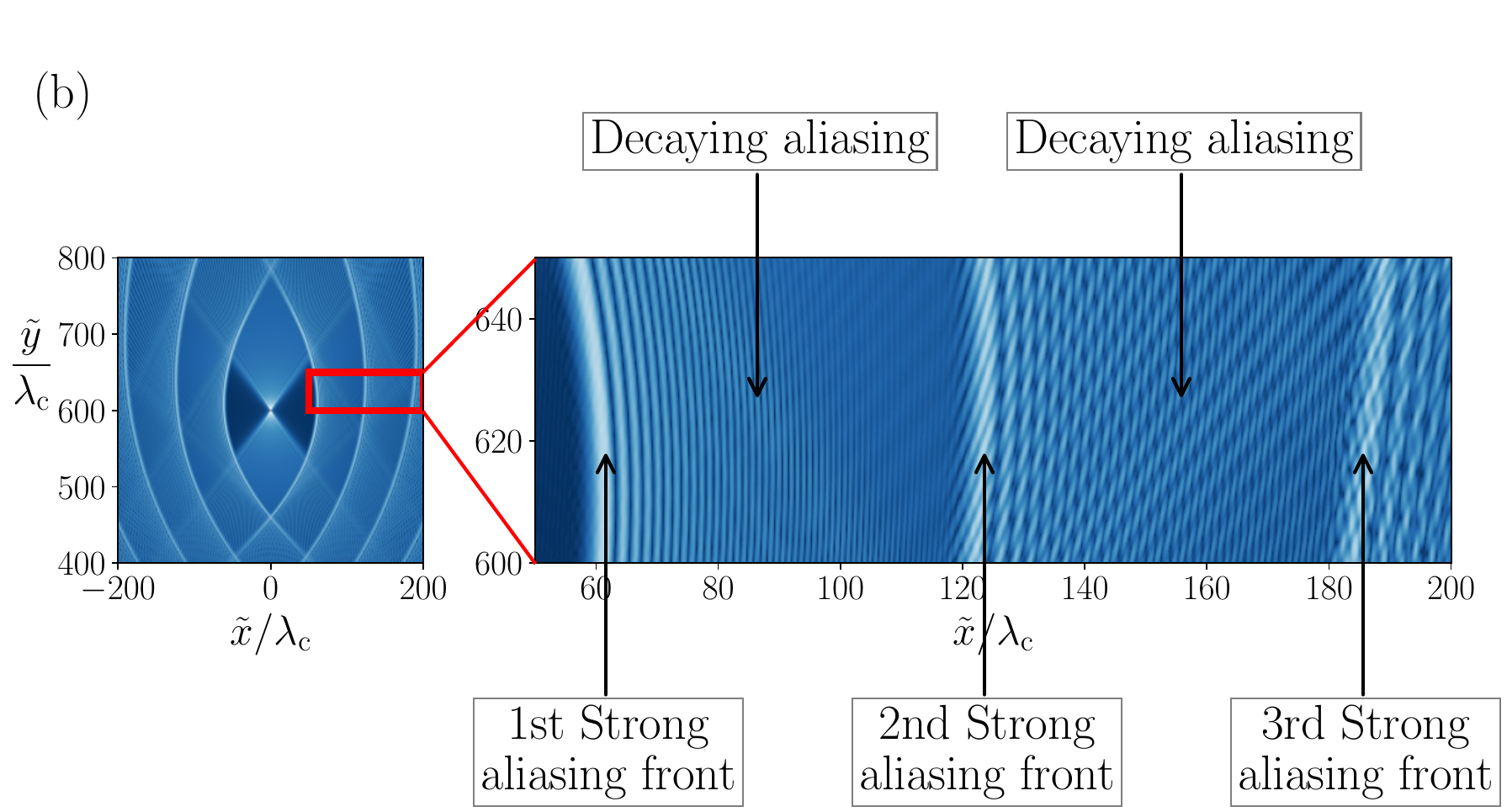}
    \caption{Representation of the spreading of the spectrum $\matchspectrum(\paramfreq \seppar \testedloc, \sourceloc)$ and its effect on the discrete-space \gls{af} displayed in Figure~\ref{fig:AF-toy}.
    (a) Spectrum for multiple locations $\testedloc$ given $\sourceloc = [0, 600]\sourcewavelen$.
    (b) Zoom in Figure~\ref{fig:AF-toy}b showing the distinct aliasing regions in the resulting discrete-space \gls{af}.}
    \label{fig:Toy-AF-Zoom}
\end{figure}

By observing the spectra in Figure~\ref{fig:toy-spectrum}, we note that the band edges exhibit sharp peaks that are a few dB higher than the bands' centers. 
When $\testedloc$ reaches a value such that one peak \rev{crosses the critical frequency} $\frac{2\pi}{\paramspacing}$, a strong aliasing front appears in $\sambifun(\testedloc, \sourceloc)$. 
As $\testedloc$ continues to move away from $\sourceloc$, the spectrum keeps spreading its energy, thereby lowering its amplitude at $\frac{2\pi}{\paramspacing}$. 
Figure~\ref{fig:Toy-AF-Zoom}a illustrates this progressive spectral spreading. 
\rev{This results in a decaying aliasing amplitude as we enter deeper into the aliased region of the \gls{af}.}
When the spectrum \rev{spreads sufficiently to cross} $\frac{4\pi}{\paramspacing}$, a second strong aliasing front appears and then decays again. 
This process repeats for all $\foldindex \geq 1$ as illustrated in Figure~\ref{fig:Toy-AF-Zoom}b, which zooms on the aliasing patterns we observed in Figure~\ref{fig:AF-toy}b.

We note that the width of these aliasing fronts reflects the width of the prominent peak in the spectra shown in Figure~\ref{fig:AF-toy}a. 
\rev{Analyzing these widths}, however, lies beyond the scope of the current study, which is focused on determining the boundaries of the aliasing-free regions.

\medskip

As discussed earlier, the spectra observed in these figures illustrate the necessity to define a soft band limit that gives a practical meaning to Lemma~\ref{lem:strict-no-aliasing}.
Since $|\matchspectrum(\paramfreq \seppar \testedloc, \sourceloc)|$ never durably reaches zero, we highlight that strictly speaking, non-zero aliasing artifacts arise for any sampling step, even finer than $\frac{\sourcewavelen}{2}$.
This has already been stated, for example, in~\cite{abhayapala_spatial_2002}.
This principle also holds in the \gls{ff} regime, where a sufficiently small sampling step that avoids grating lobes still deteriorates the theoretical cardinal sine-shaped \gls{af} into a Dirichlet kernel\footnote{A \emph{Dirichlet kernel} denotes the $\frac{\sin(N\, x)}{\sin(x)}$ function.}-shaped one. 

Consequently, Section~\ref{sec:chirp} presents our approach to define a soft band limit that enables deriving closed-form expressions of the aliasing structure in the \gls{nf} regime and remains consistent with the \gls{ff} conventions.

%% file: sections/4-Chirp.tex
\section{Chirp-based compressive spectrum modelling}
\label{sec:chirp}

To deal with the theoretical impossibility of satisfying \eqref{eq:folding-condition}, we aim to replace its right term by a softer alternative. 
Instead of an intuitive $\threshband$-tolerant band, we \rev{leverage the theoretical findings of the authors in~\cite{chassande_stationary_1998}, and follow the preliminary results we elaborated in \cite{monnoyer_chirp_2025}.}
More precisely, we exploit a chirp structure representation of the matched function $\matchfun$ to define a practical definition of the band limit. 
The approach of this section, therefore,
\begin{enumerate}
    \item provides a compressive spectral representation of $\matchspectrum(\paramfreq \seppar \testedloc, \sourceloc)$;
    \item enables \emph{closed-form expressions} of the aliasing artifact fronts in the discrete-space \gls{af}, as exemplified in the next sections; 
    \item reduces to known results in the \gls{ff} regime, hence providing a generalized framework.
\end{enumerate}

\subsection{Chirp-based analysis of aliasing artifacts}

\begin{figure*}[t]
    \centering
    \includegraphics[width=\linewidth]{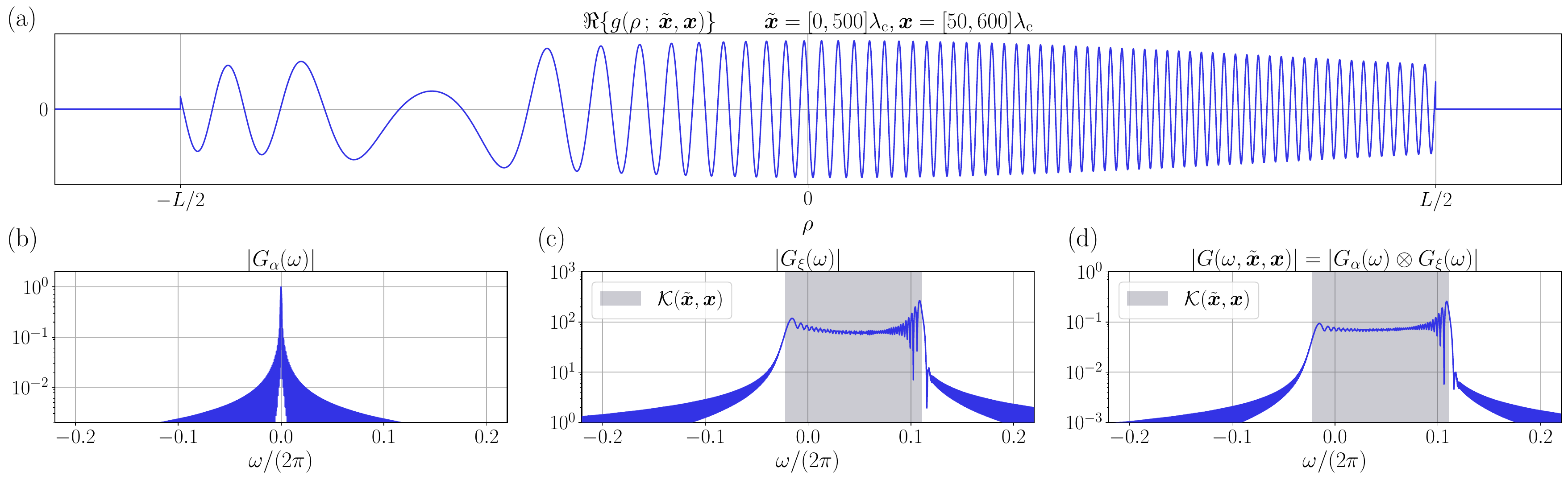}
    \caption{Representation of the chirp-structured matched signal $\matchfun(\paramvar \seppar \testedloc, \sourceloc)$ for $\sourceloc = [0, 600]\sourcewavelen$ and $\testedloc = [50, 700]\sourcewavelen$.
    (a) Real part of $\matchfun(\paramvar \seppar \testedloc, \sourceloc)$.
    (d) Corresponding spectrum, decomposed by (b) the amplitude's spectrum and (c) the phase factor's spectrum. 
    }
    \label{fig:toy-chirp}
\end{figure*}

Let us define notations to separate the matched signal by its phase and amplitude as
\begin{equation}
    \label{eq:sigprod-2}
    \matchfun(\paramvar \seppar \testedloc, \sourceloc)
    =
    \matchamp(\paramvar) \exp\big(-j \matchphase(\paramvar)\big),
\end{equation}
where we set the amplitude and phase functions
\begin{align}
    \matchamp(\paramvar) &:= 
    \amplitudefun(\paramtf(\paramvar) \seppar \testedloc) 
    \,
    \amplitudefun(\paramtf(\paramvar) \seppar \sourceloc),
    \label{eq:chirp-amplitude}\\
    \label{eq:chirp-phase}
    \matchphase(\paramvar) &:= \sourcek (\|\paramtf(\paramvar) - \sourceloc\| - \|\paramtf(\paramvar) - \testedloc\|).
\end{align}
In the above, the dependencies on $\testedloc$ and $\sourceloc$ were omitted for readability.
We interpret the non-linear phase content $\expe^{-j \matchphase(\paramvar)}$ as a spatial ``chirp", by reference to time chirp signals that are commonly used, for example, in radar systems. 
Formally, a \emph{spatial} (resp. \emph{time}) chirp is defined by a varying \emph{local wave number} (resp. \emph{instantaneous frequency}).
As notably seen in~\cite{ding_degrees_2022, ding_spatial_2024}, this local wave number is defined by the derivative $\,\dot\matchphase(\paramvar)$ of $\matchphase(\paramvar)$.
Figure~\ref{fig:toy-chirp}a plots a matched function captured in the context of the toy examples, showing its chirp nature, characterized by a slowly varying amplitude and a fast non-linear phase variation. 

Assuming a high-frequency source signal (\ie a large $\sourcek$), the phase factor $\expe^{-j \matchphase(\paramvar)}$ oscillates much faster than $\matchamp(\paramvar)$.
This allows us to neglect the latter's convolutive effect that widens the support of the matched spectrum, compared to the pure phase factor's one.
More precisely, given $\matchspectrumphi(\paramfreq)$ and $\matchspectrumabs(\paramfreq)$ denoting the \glspl{ft} of, respectively, $\expe^{-j\matchphase(\paramvar)}$ and $\matchamp(\paramvar)$, we have
\begin{equation}
    \label{eq:chirp-convolution}
    \matchspectrum(\paramfreq \seppar \testedloc, \sourceloc) = \tfrac{1}{2\pi}\matchspectrumphi(\paramfreq) \conv \matchspectrumabs(\paramfreq).
\end{equation}
\rev{The convolution in~\eqref{eq:chirp-convolution} implies that the band limit of $\matchspectrum(\paramfreq \seppar \testedloc, \sourceloc)$ equals that of $\matchspectrumphi(\paramfreq)$ plus the bandwidth of $\matchspectrumabs(\paramfreq)$.
As illustrated in Figure~\ref{fig:toy-chirp}a, the typically slow variations of $\amplitudefun(\paramvar)$ (compared to $\sourcewavelen$) yields a narrowband spectrum $\matchspectrumabs(\paramfreq)$ whose bandwidth can be neglected.}
We thus reduce our study to the content of $\matchspectrumphi(\paramfreq)$ as a good indication of $\matchspectrum(\paramfreq \seppar \testedloc, \sourceloc)$.
This assumption is further supported by Figure~\ref{fig:toy-chirp}b-d, showing the similarities between the complete spectrum and its sole phase factor counterpart.

\rev{As supported by the theoretical findings detailed in~\cite{chassande_stationary_1998}}, the set of local spatial frequencies observed by the array, defined as
\begin{equation}
    \label{eq:local-frequencies}
    \kset(\testedloc, \sourceloc) := \{\dot\matchphase(\paramvar) : \paramvar \in \paramdomain\},
\end{equation} 
constitutes a good indication of the ``most active" spatial frequency (wave number) content in the chirp's spectrum.
This has been notably formalized by the concept of ``local spatial bandwidth"~\cite{ding_degrees_2022}.
Capitalizing on this approach, Definition~\ref{def:soft-bandlimit} provides a practical expression of the spatial band limit, which matches its significant spectral content in practical cases.
Said differently, the set of local frequencies $\kset(\testedloc, \sourceloc)$ yields a compressive spectral modeling of the matched signal.
\begin{definition}[Soft Band Limit]
    \label{def:soft-bandlimit}
    The soft band limit associated with the locations $\testedloc$ and $\sourceloc$ is the maximal \emph{local} frequency of the matched signal $\matchfun(\paramvar \seppar \testedloc, \sourceloc)$.
    It is thereby defined as
    \begin{equation}
        \label{eq:chirp-bandlimit}
        \bandlim(\testedloc, \sourceloc) := \underset{\paramvar \in \paramdomain}{\max}|\dot \matchphase(\paramvar)|.
    \end{equation}
\end{definition}

Now equipped with Definition~\ref{def:soft-bandlimit}, we can substitute the impractical strict band limit $\bandlimstrict(\testedloc, \sourceloc)$ by the softer $\bandlim(\testedloc, \sourceloc)$ in Lemma~\ref{lem:strict-no-aliasing}.
Guidelines to safely design arrays straightforwardly follow in Section~\ref{sec:guidelines}.

\subsection{Aliasing-free regions and aliasing-safe operating domain}
\label{sec:guidelines}

Using the expression $\bandlim(\testedloc, \sourceloc)$ in Lemma \ref{lem:strict-no-aliasing} enables us to define a region in the discrete-space \gls{af}, $\sambifun(\testedloc, \sourceloc)$, that is free of significant aliasing. 
This leads to Definition~\ref{def:afr}.
\begin{definition}[Aliasing-Free Region]
    \label{def:afr}
    The ``\acrfull{afr}" associated with the source location $\sourceloc$ is the set of all the \emph{tested} locations $\testedloc$ that yield no aliasing when evaluating $\sambifun(\testedloc, \sourceloc)$.
    Mathematically, it is expressed as
    \begin{equation}
        \label{eq:afr}
        \afr(\sourceloc) := \{\testedloc : \bandlim(\testedloc, \sourceloc) \leq 2\pi \paramspacing^{-1}\}.
    \end{equation}
\end{definition}
The aliasing front, for instance observed in Figure~\ref{fig:AF-toy}b, is given by the boundary of $\afr(\sourceloc)$, denoted by $\partial \afr(\sourceloc)$.
We also note that $\partial \afr(\sourceloc)$ is the contour line of the function $\bandlim(\testedloc, \sourceloc)$ at the level $2\pi \paramspacing^{-1}$.

Importantly, the \gls{afr} is a \emph{location-dependent} concept since each source location $\sourceloc$ leads to a possibly different \gls{afr}.
By contrast, Definition~\ref{def:asod} proposes a global concept that only depends on the structure of the array. 
Still, both concepts depend on the sampling step $\paramspacing$, as seen in \eqref{eq:afr}.
\begin{definition}[Aliasing-Safe operating Domain]
    \label{def:asod}
    The operating domain $\sourcedomain$ is an ``\gls{asod}" if all the pairs $\testedloc, \sourceloc \in \sourcedomain$ yield no aliasing in $\sambifun(\testedloc, \sourceloc)$.
    Mathematically, this is achieved if and only if, for all $\sourceloc \in \sourcedomain$, 
    \begin{equation}
        \label{eq:asod}
        \sourcedomain \subseteq \afr(\sourceloc).
    \end{equation}
\end{definition}
We emphasize that \eqref{eq:asod} is equivalent to stating that all possible pairs of true and tested locations, $\testedloc, \sourceloc \in \sourcedomain$, are such that $\testedloc \in \afr(\sourceloc)$.
Under this condition, no grating lobe ever appears in the \gls{af} restricted to $\sourcedomain$. 
We also note that \glspl{asod} are typically not unique for a given spacing $\paramspacing$.

These definitions lead to design guidelines to avoid estimation ambiguities over a targeted operating domain $\sourcedomain$. 
\begin{enumerate}
    \item For a given array geometry $\antdomain$, obtain the soft spatial band limit $\bandlim(\testedloc, \sourceloc)$.
    \item Derive the \gls{afr} $\afr(\sourceloc)$ as a function of both $\sourceloc$ and the spacing $\paramspacing$.
    \item Using Definition~\ref{def:asod}, derive the value of $\paramspacing$ required to ensure that the domain $\sourcedomain$ is an \gls{asod}, \ie meets \eqref{eq:asod}.
\end{enumerate}

Whereas steps 1) and 2) are analytical, step 3) likely requires the development of dedicated numerical methods, extending beyond the theoretical scope of this paper.
The remainder of this paper, therefore, concentrates on the analysis of \glspl{afr}.

\subsection{Properties of aliasing-free regions}
\label{sec:afr-properties}

We can now establish properties of the \gls{afr} which strengthen our framework and are helpful in manipulating it in subsequent sections.

We begin by extending a classical \gls{ff} result to the \gls{nf} regime.
Theorem \ref{thm:lam2} exploits modelling \eqref{eq:chirp-bandlimit} to state that no aliasing can arise in the discrete-space \gls{af} as soon as the half-wavelength spacing between antennas is maintained.
In the \gls{ff} regime, which permits no range resolution from the steering signal, this principle reduces to its standard angular interpretation, guaranteeing no angular grating lobes with a half-wavelength antenna spacing.

\begin{theorem}
    \label{thm:lam2}
    Given $\antloc_i := \paramtf(i \paramspacing)$ denoting  the $i$-th antenna's locations from the grid $\antgrid$, if
    \begin{equation}
        \label{eq:thm-lam2-condition}
        \|\antloc_{i+1} - \antloc_i\| \leq \frac{\sourcewavelen}{2}
    \end{equation}
    is met for all $i\in\{1, \dots, N\text{-}1\}$,
    then the \gls{afr} is $\bb R^\ndim$ for all $\sourceloc$. 
    As a result, any $\sourcedomain \subseteq \bb R^\ndim$ is an \gls{asod}.
\end{theorem}

We emphasize the significance of this result, as Theorem~\ref{thm:lam2} extends beyond the conventional understanding of the \gls{ff} rules. 
It demonstrates that for \emph{any} set of antenna locations $\antgrid$, if at least one path connecting all these locations can be drawn such that the consecutive spacings are smaller than half the wavelength, then no aliasing occurs, neither in the \gls{nf} nor the \gls{ff} regimes.

As previously discussed, this theorem does not imply that the continuous and discrete \glspl{af}, $\ambifun(\testedloc, \sourceloc)$ and $\sambifun(\testedloc, \sourceloc)$, can be identical. 
Instead, it supports that using $\bandlim(\testedloc, \sourceloc)$ to define the soft band limit yields a \gls{nf} modeling of aliasing artifacts that matches the conventional \gls{ff} modeling of the grating lobes.

Besides its theoretical reach, the sufficient condition of Theorem~\ref{thm:lam2} can be impractical to effectively design \glspl{xla}.
By extending the array dimensions while increasing carrier frequencies, engineers face physical constraints that prevent meeting a half-wavelength spacing in some applications~\cite{lu_tutorial_2024, rodrigues_low_2020, lu_how_2021}.
This further motivates the role of our framework, focused on deriving aliasing-safe operating domains, given $\paramspacing>\sourcewavelen/2$.

\medskip

While finding practical expressions for the \glspl{afr} is sometimes challenging, we can bound it by another one that is easier to compute. 
This principle, which we exploit in the next sections, is formalized by Properties~\ref{thm:inclusion-i} and~\ref{thm:inclusion-ii}.

\begin{property}[Inclusion principle I]
    \label{thm:inclusion-i}
    Two arrays resulting from the sampling of $\paramdomain$ with respective sampling steps $\paramspacing_1 \leq \paramspacing_2$ yield respective \glspl{afr}, $\afr_1(\sourceloc)$ and $\afr_2(\sourceloc)$ that satisfy
    \begin{equation}
        \afr_2(\sourceloc) \subseteq \afr_1(\sourceloc).
    \end{equation}
    for all $\sourceloc \in \bb R^\ndim$.
\end{property}
\begin{proof}
    Since $\paramspacing_1 \leq \paramspacing_2$, the condition in \eqref{eq:afr} is softer when using the sampling step $\paramspacing_1$ rather than $\paramspacing_2$.
    Thus any $\testedloc\in\afr_2(\sourceloc)$ also meets $\testedloc\in\afr_1(\sourceloc)$.
\end{proof}

\begin{property}[Inclusion principle II]
    \label{thm:inclusion-ii}
    Given two arrays resulting from the sampling of, respectively, $\paramdomain_1$ and $\paramdomain_2$, with the same sampling step $\paramspacing$, 
    the inclusion
    \begin{equation}
        \label{eq:thm-inclusion-condition}
        \paramdomain_1 \subseteq \paramdomain_2,
    \end{equation}
    implies that their respective \glspl{afr}, $\afr_1(\sourceloc)$ and $\afr_2(\sourceloc)$ satisfy
    \begin{equation}
        \label{eq:thm-inclusion-conclusion}
        \afr_2(\sourceloc) \subseteq \afr_1(\sourceloc),
    \end{equation}
    for all $\sourceloc \in \bb R^\ndim$.
\end{property}
\begin{proof}
    Given \eqref{eq:thm-inclusion-condition}, $\bandlim_2(\testedloc, \sourceloc)$ is the solution of a \emph{relaxed} version of the maximization problem \eqref{eq:chirp-bandlimit} that yields $\bandlim_1(\testedloc, \sourceloc)$.
    Therefore, for all $\testedloc, \sourceloc$, we have $
        \bandlim_2(\testedloc, \sourceloc) \geq \bandlim_1(\testedloc, \sourceloc).$
    Consequently 
    \begin{equation}
        \bandlim_2(\testedloc, \sourceloc) \leq 2\pi \paramspacing^{-1}
        \Rightarrow
        \bandlim_1(\testedloc, \sourceloc) \leq 2\pi \paramspacing^{-1}.
    \end{equation}
    This directly proves \eqref{eq:thm-inclusion-conclusion} by definition of the \gls{afr}.
\end{proof}

While Property~\ref{thm:inclusion-i} states that densifying the array with extra antennas improves (widens) the \glspl{afr}, Property~\ref{thm:inclusion-ii} implies that extending the size of the array deteriorates (narrows) the \glspl{afr}.
Conversely, we deduce that truncating an antenna array can only \emph{widen} its \glspl{afr}.
This possibly counterintuitive conclusion stems from the fact that aliasing artifacts alter the \gls{af} as soon as \emph{one} local spatial frequency is greater than $2\pi\paramspacing^{-1}$.
Extending the array provides more opportunities for this condition to be met in at least one location in the array.
This contrasts with the natural intuition from the \gls{ff} regime, where the matched signal maintains a constant local spatial frequency along the array.
In that case, extending the array has no consequence on the aliasing, but solely improves the system's resolution.
Similarly, in the \gls{nf} regime, narrower arrays are not universally preferable since large arrays also offer enhanced spatial resolution for beam pointing.
A fundamental trade-off therefore follows between the resolution and the ambiguities caused by aliasing artifacts. 
Additionally, while wider arrays exhibit aliasing artifacts closer to $\sourceloc$ in the \gls{af}, their amplitude, which remains to be evaluated, might be lower.
This motivates future studies addressing aliasing not only by presence, but also by amplitude.

Finally, while the above properties affect \glspl{afr}, Property~\ref{thm:inclusion-iii} extends their conclusions to \glspl{asod}.

\begin{property}[Inclusion principle III]
    \label{thm:inclusion-iii}
    When two arrays are providing \glspl{afr} such that
    \begin{equation}
        \label{eq:TODO}
        \afr_2(\sourceloc) \subseteq \afr_1(\sourceloc)
    \end{equation}
    holds for all $\sourceloc \in \bb R^\ndim$, 
    any operating domain $\sourcedomain$ that is an \gls{asod} for the array ``2" is also an \gls{asod} for the array ``1".
\end{property}
\begin{proof}
    Combining the definition~\eqref{eq:asod} with \eqref{eq:TODO}, we obtain that if $\sourcedomain$ is an \gls{asod} for the array ``2", then for all $\sourceloc\in\sourcedomain$, we have $\sourcedomain\subseteq\afr_2(\sourceloc)\subseteq \afr_1(\sourceloc)$.
\end{proof}

The next two sections present applications of the different steps of our methodology to two canonical array topologies, \ie the \gls{ula} and the \gls{uca}.

%% file: sections/5-ULA.tex
The next two sections present applications of the different steps of our methodology to two canonical array topologies, \ie the \gls{ula} and the \gls{uca}.

\section{Uniform linear array the NF regime}
\label{sec:ula}

\begin{figure}[t]
    \begin{center}
    \includegraphics[width=0.85\linewidth]{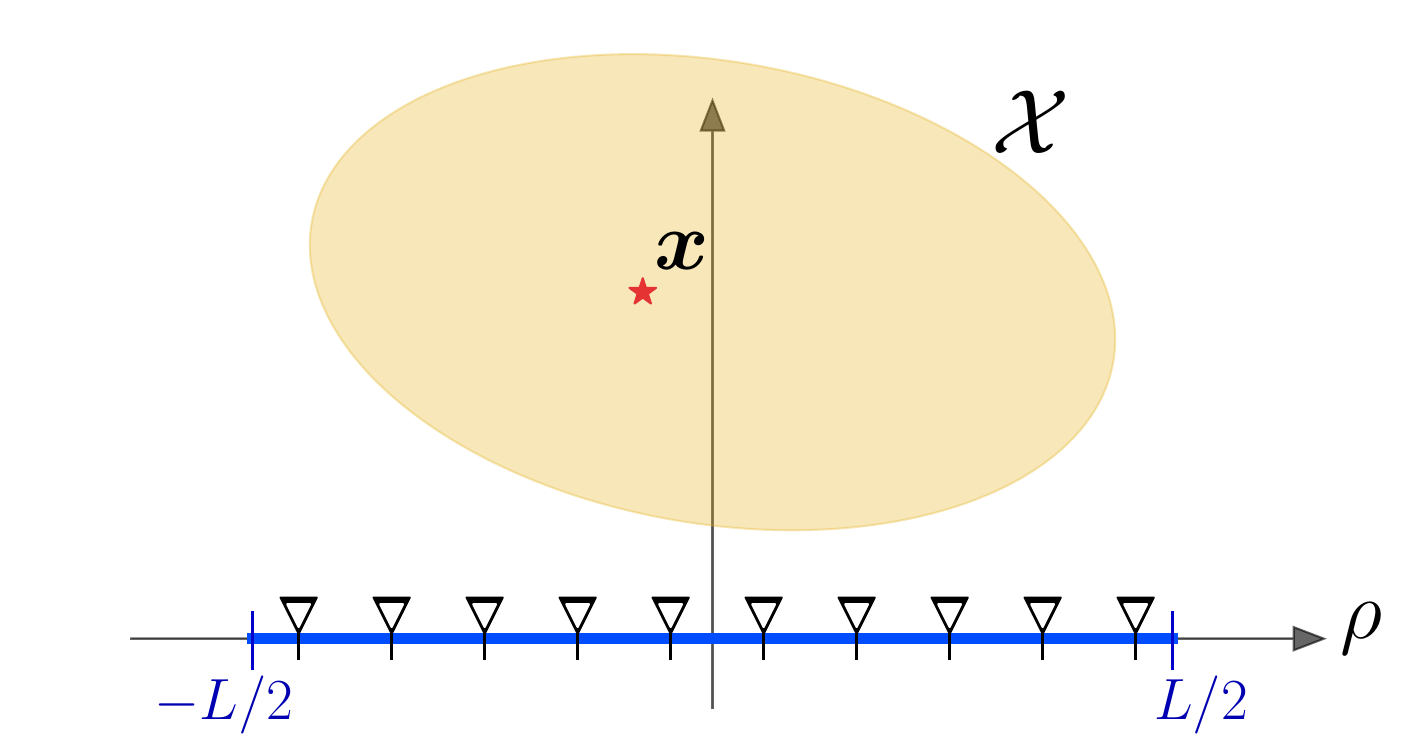}
    \end{center}
    \caption{Schematic representation of a \gls{ula} operating in the \gls{nf} regime.}
    \label{fig:ULAscenario}
    \vspace{-5mm}
\end{figure}

Using the theoretical framework presented in the previous sections, we further develop the specific case of a \gls{ula}.
Without loss of generality, we consider the zero-centered and horizontal \gls{ula}, similar to the toy example used in Section~\ref {sec:framework}. 
Its geometry, shown in Figure~\ref{fig:ULAscenario}, is trivially described with $\paramtf(\paramvar) = [\paramvar, 0]^\top$ for $\paramvar \in \paramdomain_\lenula:=[-\lenula/2, \lenula/2]$, where $\lenula$ denotes its length.

\begin{figure*}[t]
    \centering
    \includegraphics[width=\linewidth]{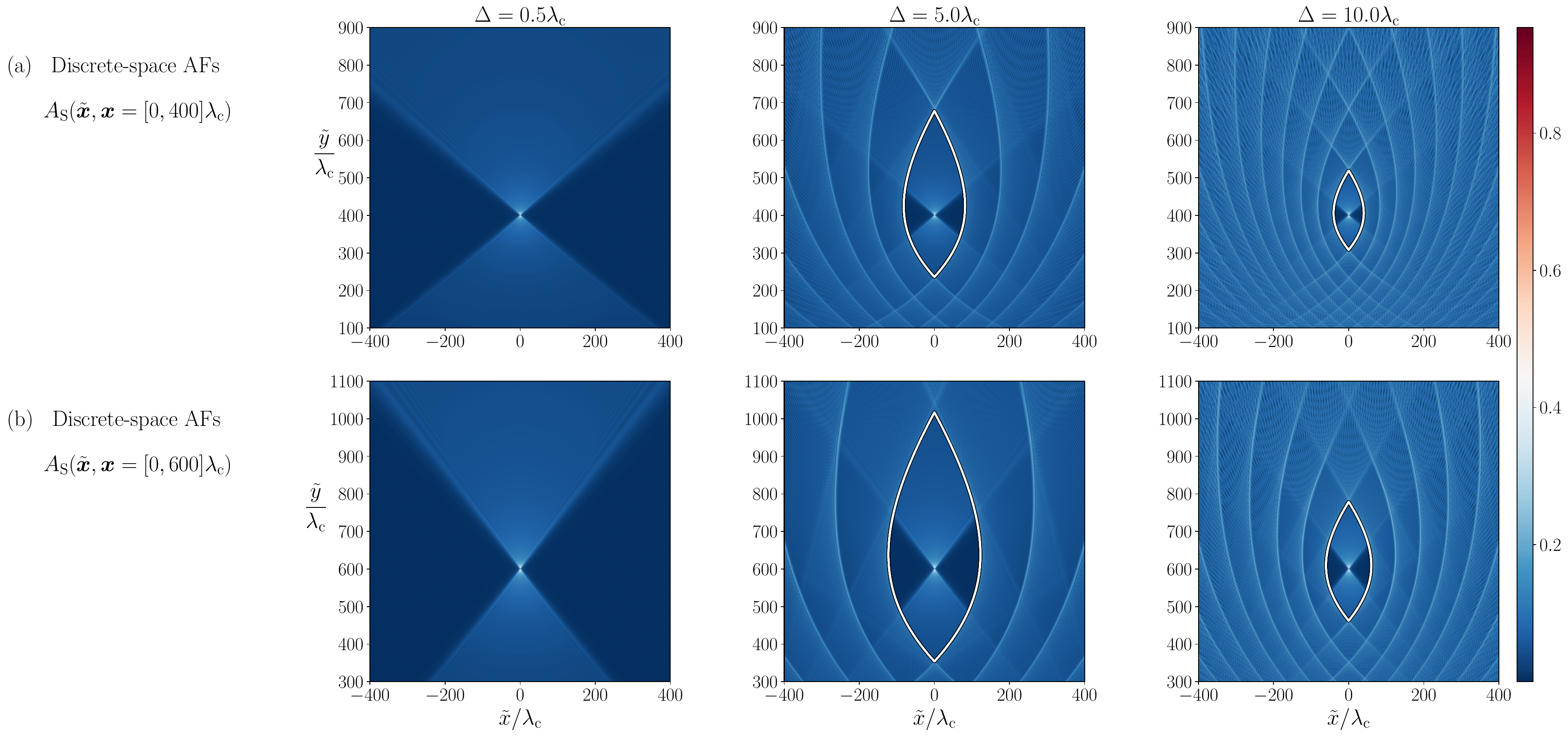}
    \caption{Discrete-space \glspl{af} obtained for a horizontal \gls{ula} of length $\lenula = 1000\sourcewavelen$.
    (a) with $\sourceloc=[0, 400] \sourcewavelen$ and (b) with $\sourceloc=[0, 600] \sourcewavelen$.
    The white eye depicts the contour line where $\bandlim_\infty(\testedloc, \sourceloc) = 2\pi \paramspacing^{-1}$, \ie $\partial \afr_\infty(\sourceloc)$, increasing in size as the distance $\yloc$ increases and shrinking as $\paramspacing$ increases.}
    \label{fig:AF-ULA-Vertical}
\end{figure*}

Following the steps given in Section~\ref{sec:guidelines}, this section derives the expression of the soft band limit, which we denote here by $\bandlim_\lenula(\testedloc, \sourceloc)$, and then deduces the structure of the corresponding \gls{afr}, denoted by $\afr_\lenula(\sourceloc)$.
Whereas our preliminary work in~\cite{monnoyer_chirp_2025} proposed an approximation of $\bandlim_\lenula(\testedloc, \sourceloc)$, whose validity was limited by the scope of the Fresnel approximation, the current paper derives its exact expression.

\rev{
Property~\ref{thm:inclusion-ii} motivates the study of the aliasing structure of the \emph{infinite-length} \gls{ula} as a \emph{worst-case reference}.
Indeed, this property guarantees 
\begin{equation}
    \label{eq:ula-inclusion}
    \afr_\infty(\sourceloc) \subseteq \afr_\lenula(\sourceloc),
\end{equation}
for all finite length $\lenula$. 
As a result, $\afr_\infty(\sourceloc)$ provides a pessimistic representation of the \gls{afr}. 
Moreover, since $\paramdomain_\infty = \bb R$, the resulting unconstrained computation \eqref{eq:chirp-bandlimit} of the soft band limit is simplified. 

For these reasons, we begin this section by applying our methodology to the infinite \gls{ula}.
We also investigate under which conditions the inclusion \eqref{eq:ula-inclusion} is tight with equality.
Those are the cases where finite-size \glspl{ula} behave as infinite in terms of grating lobes. 
Next, we provide the general expression of $\bandlim_\lenula(\testedloc, \sourceloc)$, for $\lenula < \infty$, and study the geometry of the resulting \gls{afr}.
}

\rev{
\subsection{Infinite-length ULA}
\label{sec:ula-infinite}
}

Let us now derive the closed-form expression of the infinite-length \gls{ula}'s band limit, $\bandlim_\infty(\testedloc, \sourceloc)$ for all 2D locations,
\begin{equation}
    \sourceloc = [\xloc, \yloc]^\top \quad \text{and} \quad \testedloc = [\xtest, \ytest]^\top,
\end{equation}
in the upper half-plane ($\yloc>0$ and $\ytest>0$).
By symmetry, our conclusions directly extend to the other half-plane. 

This band limit's expression is formalized in Theorem~\ref{thm:bandlimit-ula}, which is proven in Appendix~\ref{app:ula}.
Interestingly, it is entirely driven by two scalar variables
\begin{equation}
    \label{eq:ula-bandlim-ratio}
    \yratio := \bigg(\frac{\ytest}{\yloc}\bigg)^\frac{2}{3}
    \quad  \text{and} \quad
    \xratio := \frac{\xtest-\xloc}{\yloc},
\end{equation}
reducing it from four degrees of freedom to two.
To simplify notations, we also set $\xyratio := \displaystyle\frac{\xratio}{\yratio^2 -1}$.

\begin{theorem}[Band limit of an infinite-length ULA]
    The infinite-length \gls{ula} yields a soft band-limit (Definition~\ref{def:soft-bandlimit}) that is described by the closed-form expression
    \label{thm:bandlimit-ula}
    \begin{equation}
        \label{eq:thm-bandlimit-ula}
        \bandlim_\infty(\testedloc, \sourceloc) = 
        \sourcek
        \frac{|(\yratio - 1) \prootulan + \xratio|}{\sqrt{\yratio(\prootulan^2 + 1)}},
    \end{equation}
    where
    \begin{equation}
        \label{eq:thm-bandlimit-ula-root}
        \prootulan := 
        \sign(\xyratio) \yratio \sqrt{(\yratio+1)^{-1} + \xyratio^2} - \xyratio.
    \end{equation}
\end{theorem}

Inspecting the expression \eqref{eq:thm-bandlimit-ula} for specific locations $\testedloc$ that are aligned with the source location $\sourceloc$ yields simplifications formulated in Corollaries~\ref{cor:bandlimit-ula-horizontal} and~\ref{cor:bandlimit-ula-vertical}, proven in Appendix~\ref{app:ula-cor}. 
These particular cases are similar to the spatial bandwidth expressions obtained in~\cite{ding_degrees_2022}.

\begin{corollary}
    \label{cor:bandlimit-ula-horizontal}
    When $\testedloc$ and $\sourceloc$ are horizontally aligned, such that $\yloc = \ytest$, the band limit expression \eqref{eq:thm-bandlimit-ula} is reduced to
    \begin{equation}
        \label{eq:thm-bandlimit-ula-u1}
        \bandlim_\infty(\testedloc, \sourceloc) = 
        \sourcek
        \frac{|\xratio|}{\sqrt{\xratio^2/4 +1}}.
    \end{equation}
\end{corollary}
\begin{corollary}
    \label{cor:bandlimit-ula-vertical}
    When $\testedloc$ and $\sourceloc$ are vertically aligned, such that $\xloc = \xtest$, the band limit expression \eqref{eq:thm-bandlimit-ula} is reduced to
    \begin{equation}
        \label{eq:thm-bandlimit-ula-v0}
        \bandlim_\infty(\testedloc, \sourceloc) = 
        \sourcek
        \frac{|\yratio -1|}{\sqrt{1 + \yratio + \yratio^2}}.
    \end{equation}
\end{corollary}

\medskip

Using Theorem~\ref{thm:bandlimit-ula}, we obtain the \gls{afr} of the infinite-length \gls{ula},
\begin{equation}
    \label{eq:afr-ula-infty}
    \afr_\infty(\sourceloc) = \{\testedloc : \bandlim_\infty(\testedloc, \sourceloc) \leq 2\pi \paramspacing^{-1}\}.
\end{equation}

To illustrate this region, Figure~\ref{fig:AF-ULA-Vertical} presents numerically computed \glspl{af} for long \glspl{ula} with $\lenula = 1000 \sourcewavelen$, considering multiple antenna spacings $\paramspacing$ and two distinct source locations $\sourceloc$.
Although the simulated arrays have a finite length, the results in Figure~\ref{fig:AF-ULA-Vertical} are generated in a regime that exactly yields $\afr_\lenula(\sourceloc) = \afr_\infty(\sourceloc)$, as further explained in Section~\ref{sec:ula-finite}. 
We can therefore assume that these long \glspl{ula} behave as effectively infinite with respect to the grating lobes. 

In that figure, the white eye-shaped contours depict the boundaries $\partial \afr_\infty(\sourceloc)$, computed using the expression \eqref{eq:thm-bandlimit-ula} in \eqref{eq:afr-ula-infty}.
Their apparent correspondence with the actual aliasing front observed in the \glspl{af} illustrates that our methodology, based on \emph{local} spatial frequency content of the matched signal, correctly captures the aliasing structure appearing in such \glspl{af}.

\medskip

We now elaborate on the properties of $\afr_\infty(\sourceloc)$. 
First, we note that $\bandlim_\infty(\testedloc, \sourceloc)$ only depends on $\yratio$ and $\xratio$.
Their definition in \eqref{eq:ula-bandlim-ratio} translates an $\xloc$-shift and $\yloc$-scaling invariance. This effect is detailed in Property~\ref{prop:ula-inv}, which we prove in Appendix~\ref{app:ula-afr-prop}.
\begin{property}[Invariance]
    \label{prop:ula-inv}
    Given a spacing $\paramspacing$, the \gls{afr} around the true location $\sourceloc$ is identical to the ``ULA-eye", defined as
    \begin{equation}
        \label{eq:ula-eye}
        \ulaeye := \afr_\infty([0, 1]^\top), 
    \end{equation}
    up to a translation---centering it in $\sourceloc$--- and a rescaling by a factor $\yloc$.
    That is,
    \begin{equation}
        \label{eq:ula-invariance}
        \afr_\infty(\sourceloc) = \yloc \cdot \ulaeye + 
        \left[\begin{array}{c}
             \xloc \\
             0 
        \end{array}\right].
    \end{equation}
\end{property}

Property~\ref{prop:ula-inv} is an important result enabling us to restrict, without loss of generality, the study of $\afr_\infty(\sourceloc)$ to the one of the ``ULA eye" $\ulaeye$, whose sole remaining dependency is with the antenna spacing $\paramspacing$.
Subsequently, we describe the aperture of the eye as a function of $\paramspacing$ in Properties~\ref{prop:ula-symmetry} and~\ref{prop:ula-eye-aperture} (also proven in Appendix~\ref{app:ula-afr-prop}).

\begin{figure}[t]
    \centering
    \includegraphics[width=\linewidth]{figures/AF/ULA-eyes.pdf}
    \caption{Geometry of the ULA-eyes.
    (a) representation of its width $\eyewidth$, upper and lower aperture $\eyetop$, $\eyebot$.
    (b)-(c) Examples of ULA-eyes for multiple antenna spacing $\paramspacing$.}
    \label{fig:ULA-eye}
\end{figure}

\begin{property}[Symmetry]
    \label{prop:ula-symmetry}
    If $\testedloc \in \ulaeye$, then its $\mathrm{x}$-symmetric location $[-\xtest, \ytest]^\top \in \ulaeye$.
\end{property}

\begin{property}[Eye's aperture]
    \label{prop:ula-eye-aperture}
    The vertical and horizontal aperture of the ULA eye, characterized by $\eyewidth$, $\eyetop$, and $\eyebot$ as depicted in Figure~\ref{fig:ULA-eye}a, are given by
    \begin{align}
        \label{eq:eyewidth}
        \eyewidth &= \frac{2}{\sqrt{4\spacingratio^2 -1}}\\
        \label{eq:eyetop}
        \eyetop   &= 
        \Bigg\{
        \begin{array}{ll}
            \Big[\tfrac{2\spacingratio^2 +1 + \sqrt{12\spacingratio^2 -3}}{2(\spacingratio^2-1)}\Big]^\frac{3}{2}
            & \text{if } \spacingratio > 1 \\
            \infty & \text{otherwise}
        \end{array},
        \\
        \label{eq:eyebot}
        \eyebot   &= 
        \Bigg\{
        \begin{array}{ll}
            \Big[\tfrac{2\spacingratio^2 +1 - \sqrt{12\spacingratio^2 -3}}{2(\spacingratio^2-1)}\Big]^\frac{3}{2}
            & \text{if } \spacingratio > 1 \\
            0 & \text{otherwise}
        \end{array},
    \end{align}
    where
    \begin{equation}
        \spacingratio = \frac{\paramspacing}{\sourcewavelen}.
    \end{equation}
\end{property}

Figures \ref{fig:ULA-eye}b and~\ref{fig:ULA-eye}c show the eyes for multiple spacings $\paramspacing$.
In accordance with Property~\ref{thm:inclusion-i}, increasing this spacing shrinks the eye.
Smaller eyes exhibit near vertical-symmetry, while larger ones elongate upward, \rev{until} $\sourcewavelen/2 < \paramspacing \leq \sourcewavelen$.
In that case, expressions \eqref{eq:eyetop} and \eqref{eq:eyebot} saturate, causing the eye to degenerate.
The degenerate eyes (Fig.~\ref{fig:ULA-eye}c) never close vertically and are squeezed on the bottom. 
When $\paramspacing \leq \sourcewavelen/2$, the eye no longer has boundaries because, as stated in Theorem~\ref{thm:lam2}, we have $\ulaeye = \bb R^2$.

\begin{figure}[t]
    \centering
    \includegraphics[width=\linewidth]{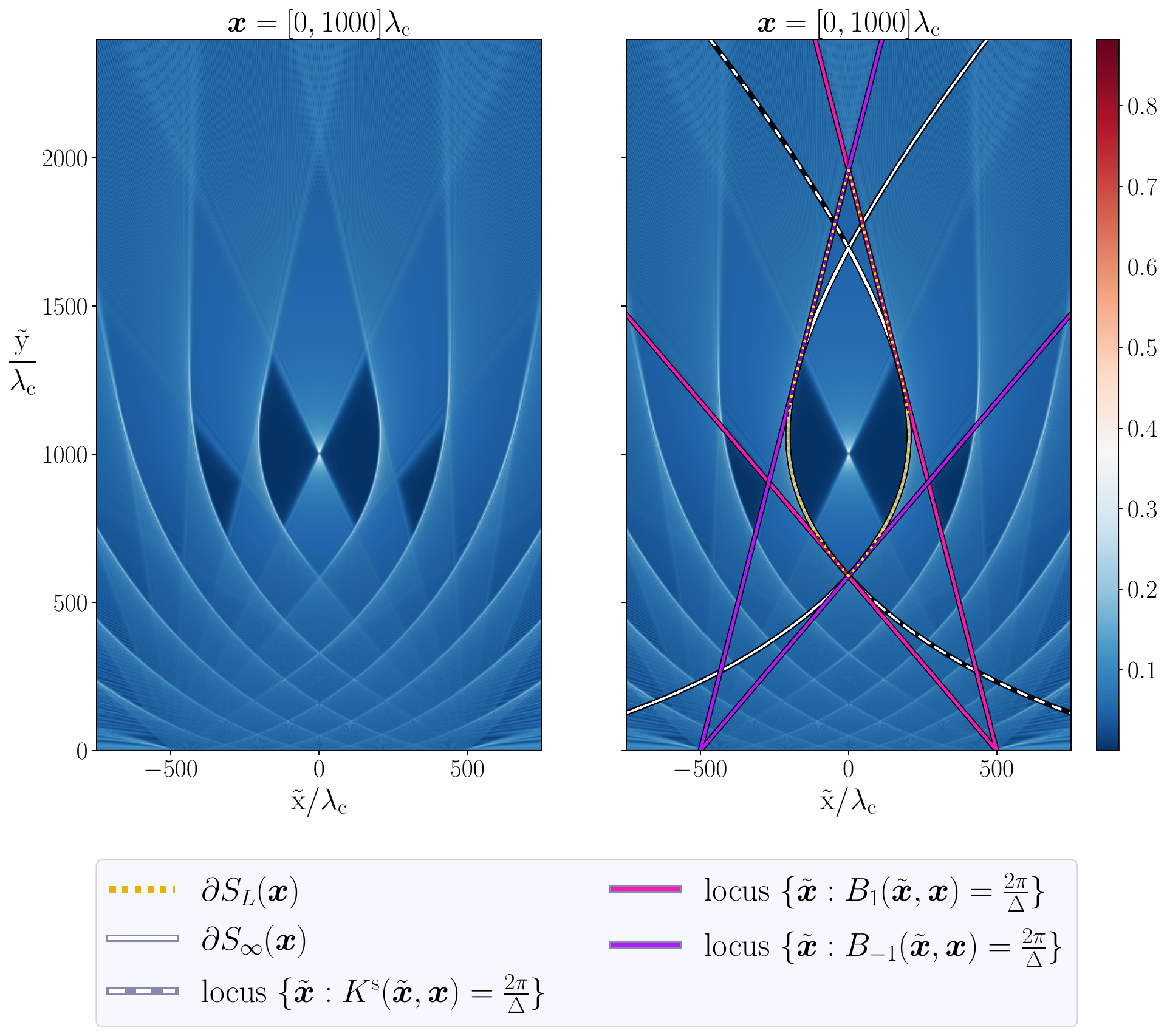}
    \caption{Ambiguity function $\sambifun(\testedloc, \sourceloc = [0, 1000]\sourcewavelen)$ for a \gls{ula} with length $\lenula= 1000\sourcewavelen$ and antenna spacing $\paramspacing = 5\sourcewavelen$; \textbf{(left)} without annotations and \textbf{(right)} with annotations corresponding to the three regimes of the band limit.}
    \label{fig:AF-ULA-annotations}
\end{figure}

\rev{

\subsection{Finite-length ULA}
\label{sec:ula-finite}

\begin{figure*}[t]
    \centering
    \includegraphics[width=\linewidth]{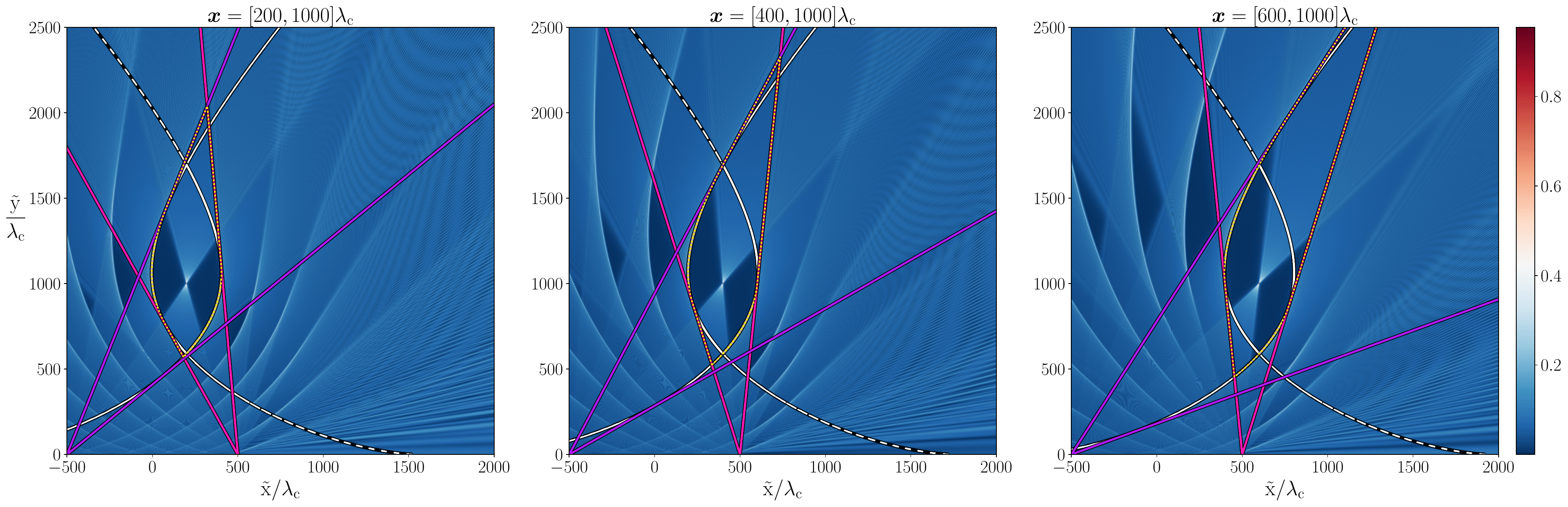}
    \caption{Ambiguity functions for a \gls{ula} with length $\lenula= 1000\sourcewavelen$ and antenna spacing $\paramspacing = 5\sourcewavelen$ for multiple source locations $\sourceloc$ that yield distinct \gls{afr} shapes, combining the three regimes explained in Section~\ref{sec:ula-finite}. The legend is identical to Figure~\ref{fig:AF-ULA-annotations}.}
    \label{fig:AF-ULA-xshift}
\end{figure*}

When the length $\lenula$ is finite, the expressions for the soft band limit and the resulting \gls{afr} become more sophisticated.
Theorem~\ref{thm:bandlinit-ula-L}, proven in Appendix~\ref{proof:thm-ula-L}, provides the composite expression of $\bandlim_\lenula(\testedloc, \sourceloc)$, which exhibits three regimes that we explain in this section.
\renewcommand{\arraystretch}{1.2}
\begin{theorem}[Band limit of an finite-length ULA]
    \label{thm:bandlinit-ula-L}
    The \gls{ula} of length $\lenula$ yields a soft band-limit (Definition~\ref{def:soft-bandlimit}) that is described by the closed-form expression
    \begin{multline}
        \label{eq:thm2-bandlim-ula}
        \bandlim_\lenula (\testedloc, \sourceloc) 
        = \\
        \left\{
        \begin{array}{ll}
            \bandlim_\infty(\testedloc, \sourceloc) 
                & \text{if } |\yloc \prootulan + \xloc| \leq \frac{\lenula}{2} \\
            \max\{\bandlimplus(\testedloc, \sourceloc), \bandlimsec(\testedloc, \sourceloc)\}
                & \text{if } |\yloc \prootulan + \xloc| > \frac{\lenula}{2}\\
                & \text{ and } |\yloc \prootulansec + \xloc| \leq \frac{\lenula}{2}\\
            \max\{\bandlimplus(\testedloc, \sourceloc), \bandlimminus(\testedloc, \sourceloc)\}
                & \text{otherwise} 
        \end{array}
        \right.,
    \end{multline}
    where
    \begin{align}
        \label{eq:bandlimpm}
        \bandlimpm(\testedloc, \sourceloc) 
        &=
        \sourcek \big[
        \cos\big(\theta_q(\testedloc)\big)
        -
        \cos\big(\theta_q(\sourceloc)\big)
        \big], \\
        \label{eq:bandlimsec}
        \bandlimsec(\testedloc, \sourceloc) 
        &= 
        \sourcek
        \frac{|(\yratio - 1) \prootulansec + \xratio|}{\sqrt{\yratio((\prootulansec)^2 + 1)}},
    \end{align}
    and where, for $q\in\{-1, 1\}$,
    \begin{align} 
        \label{eq:ula-angle}
        \theta_q(\sourceloc) &:= \angle\big(
        \sourceloc - [q\:\sign(\xyratio)\tfrac{\lenula}{2}\,, \, 0]^\top
        \big),\\
        \label{eq:thm2-bandlimit-ula-root-secondary}
        \prootulansec &:= 
        -\sign(\xyratio) \yratio \sqrt{(\yratio+1)^{-1} + \xyratio^2} - \xyratio.
    \end{align}

\end{theorem}
}

\begin{figure*}[t]
    \centering
    \includegraphics[width=\linewidth]{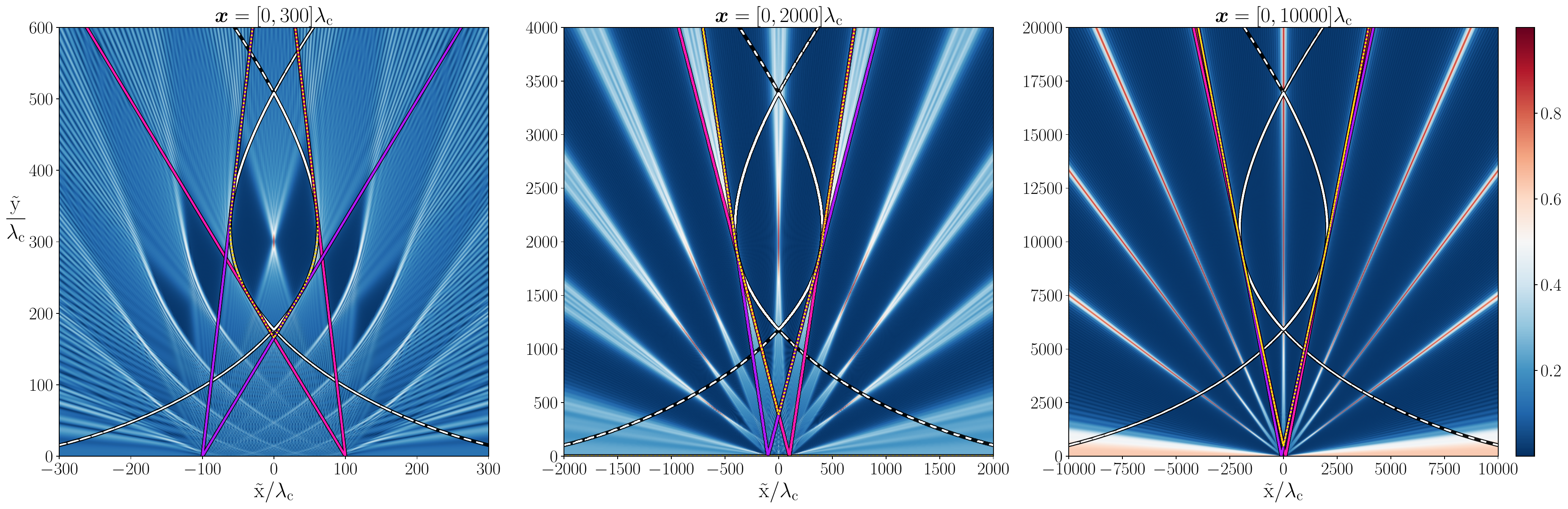}
    \caption{Ambiguity functions for a \gls{ula} with length $\lenula= 200\sourcewavelen$ and antenna spacing $\paramspacing = 5\sourcewavelen$ for multiple source locations $\sourceloc$ that progressively tends towards the \gls{ff} conditions. 
    The location $\sourceloc = [0, 10000]\sourcewavelen$ is in the \gls{ff} according to the Fraunhofer criterion.
    The legend is identical to Figure~\ref{fig:AF-ULA-annotations}.}
    \label{fig:AF-ULA-NF-FF}
\end{figure*}

The first regime of \eqref{eq:thm2-bandlim-ula} highlights the equivalence $\bandlim_\lenula(\testedloc, \sourceloc) = \bandlim_\infty(\testedloc, \sourceloc)$ for all pairs $\testedloc, \sourceloc$ such that
\begin{equation}
    \label{eq:ula-maximizer-ok}
    |\yloc \prootulan + \xloc| \leq \tfrac{\lenula}{2}.
\end{equation} 
This can be explained since by construction of Theorem~\ref{thm:bandlimit-ula}'s proof (see Appendix~\ref{app:ula}), the variable $\prootulan$ is linked to the value of $\paramvar$ that sees the highest local frequency, through 
\begin{equation}
    \label{eq:bandlimit-ula-maximizer}
    \yloc\prootulan + \xloc = \argmax_{\paramvar\in\bb Z}|\dot\matchphase(\paramvar)|.
\end{equation}
Therefore, under \eqref{eq:ula-maximizer-ok}, the constraint $\paramvar \in [-\lenula/2, \lenula/2]$ does not exclude the global maximizer of the unconstrained maximization that underpins $\bandlim_\infty(\testedloc, \sourceloc)$. 
\rev{Corollary~\ref{cor:afr-ula-inf-noninf} directly follows.

\begin{corollary}
    \label{cor:afr-ula-inf-noninf}
    Consider an infinite-length \gls{ula} and its truncation of length $\lenula$. For a given location $\sourceloc$, if for all $\testedloc \in \afr_\infty(\sourceloc)$ the condition \eqref{eq:ula-maximizer-ok} holds, then
    \begin{equation}
    \afr_\lenula(\sourceloc) = \afr_\infty(\sourceloc).
    \end{equation}
\end{corollary}
\begin{proof}
    This is a direct consequence of Theorem~\ref{thm:bandlinit-ula-L}.
\end{proof}
}
Despite the non-trivial expression of $\prootulan$ (which depends on both $\sourceloc$ and $\testedloc$), we can intuitively anticipate that \eqref{eq:ula-maximizer-ok} is typically satisfied when $\sourceloc$ and $\testedloc$ are sufficiently close to each other and to the array's center, compared to the length $\lenula$. 
Such close-by locations see the array as effectively infinite.

\rev{
Conversely, when this condition is not met, the band limit expression is mainly driven by its third regime, \ie $\bandlim_\lenula (\testedloc, \sourceloc) = \max\{\bandlimplus(\testedloc, \sourceloc), \bandlimminus(\testedloc, \sourceloc)\}$.
The second regime describes the transition between the two others.
In the third regime, the highest local frequency occurs at one of the array edges, revealing, in \eqref{eq:bandlimpm}, the well-known difference of directional cosines with angles taken relative to the given edge.
}

\rev{
Expression \eqref{eq:thm2-bandlim-ula} can now be exploited to compute $\afr_\lenula(\sourceloc)$. 
Unlike Figure~\ref{fig:AF-ULA-Vertical}, which focused on the ``near-infinite array" regime, Figure~\ref{fig:AF-ULA-annotations} shows results with a further source location, causing $\afr_\lenula(\sourceloc) \neq \afr_\infty(\sourceloc)$. 
The bottom part of the boundary $\partial \afr_\lenula(\sourceloc)$ matches $\partial \afr_\infty(\sourceloc)$ as it corresponds to locations $\testedloc$ that meets condition \eqref{eq:ula-maximizer-ok}.
The top part, however, is attained in the third regime of the band limit, causing $\partial \afr_\lenula(\sourceloc)$ to instead follow the directional cosine lines (in pink).

Figure~\ref{fig:AF-ULA-xshift} shows similarly annotated \glspl{af} for multiple source locations $\sourceloc$, illustrating how the corresponding \glspl{afr} vary in shape while consistently adhering to the lines marking the regimes we detailed in this section.

}

\rev{

\subsection{Connections with the far-field regime}
\label{sec:ula-ff}

To conclude our analysis of \glspl{ula}, we examine the behavior of our \gls{afr} expression as the scenario approaches the \gls{ff} regime. 
Figure~\ref{fig:AF-ULA-NF-FF} presents results similar to those in Figure~\ref{fig:AF-ULA-xshift}, with the source locations $\sourceloc$ that are progressively moved further from a \gls{ula} of length $200\sourcewavelen$.
The right-side subfigure notably shows the case where $\sourceloc$ is in the \gls{ff} as it reaches the Fraunhofer distance.
As the distances become much larger than the array length $\lenula$, two effects emerge: (i) the ``directionnal cosine" regime becomes the sole active regime in the boundary $\partial \afr_\lenula(\sourceloc)$, and (ii) the cosines in \eqref{eq:bandlimpm} measured from both array edges converge.
The combination of these two effects connects our model with the seminal angular repetition, characteristic of the \gls{ff} regime.
}

%% file: sections/6-UCA.tex
\begin{figure}[tb]
    \begin{center}
    \includegraphics[width=0.95\linewidth]{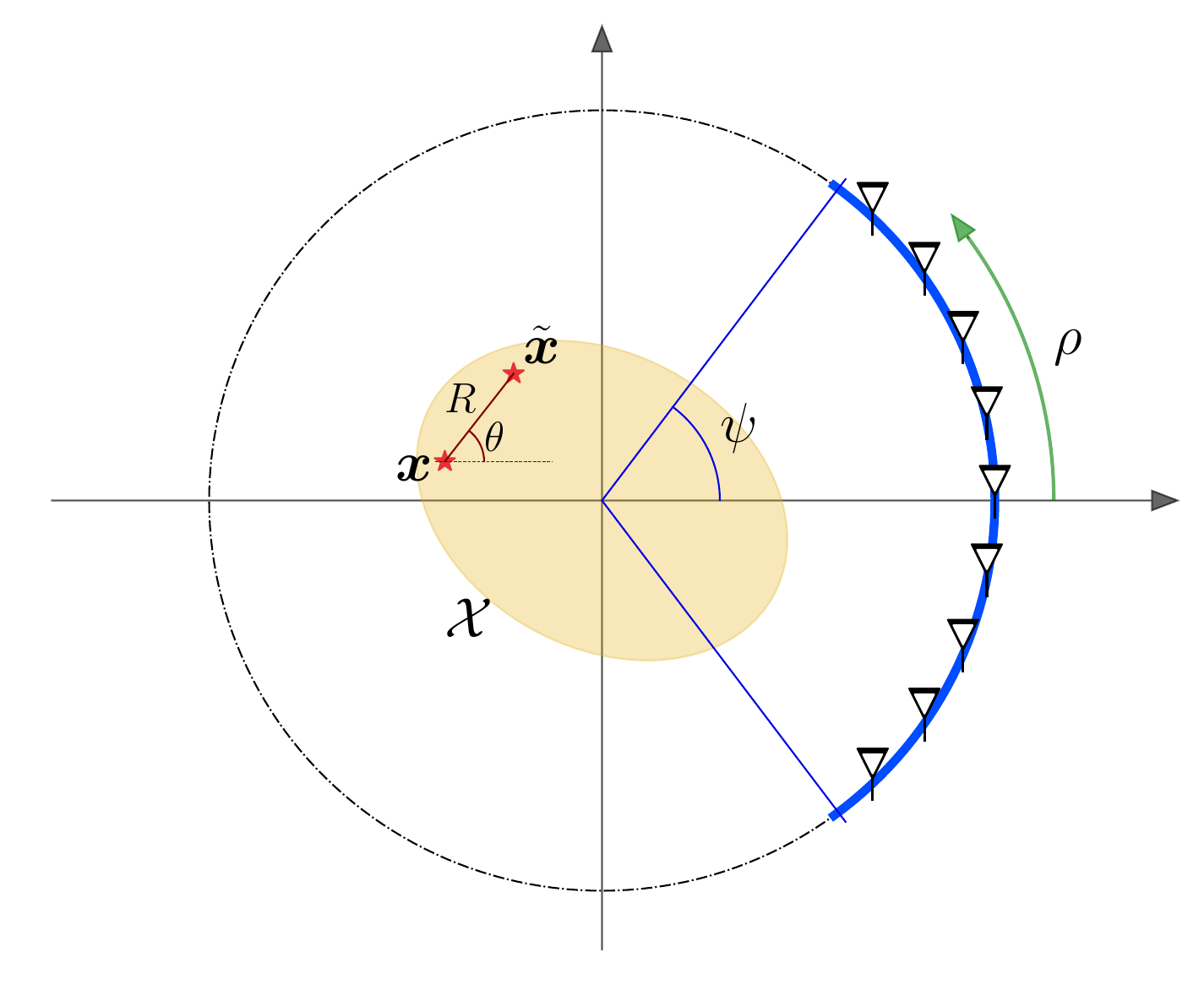}
    \end{center}
    \caption{Schematic representation of a \gls{uca}}
    \label{fig:uca-scenario}
\end{figure}

\section{Distributed Uniform Circular Array}
\label{sec:uca}

\begin{figure}[!tb]
    \centering
    \includegraphics[width=\linewidth]{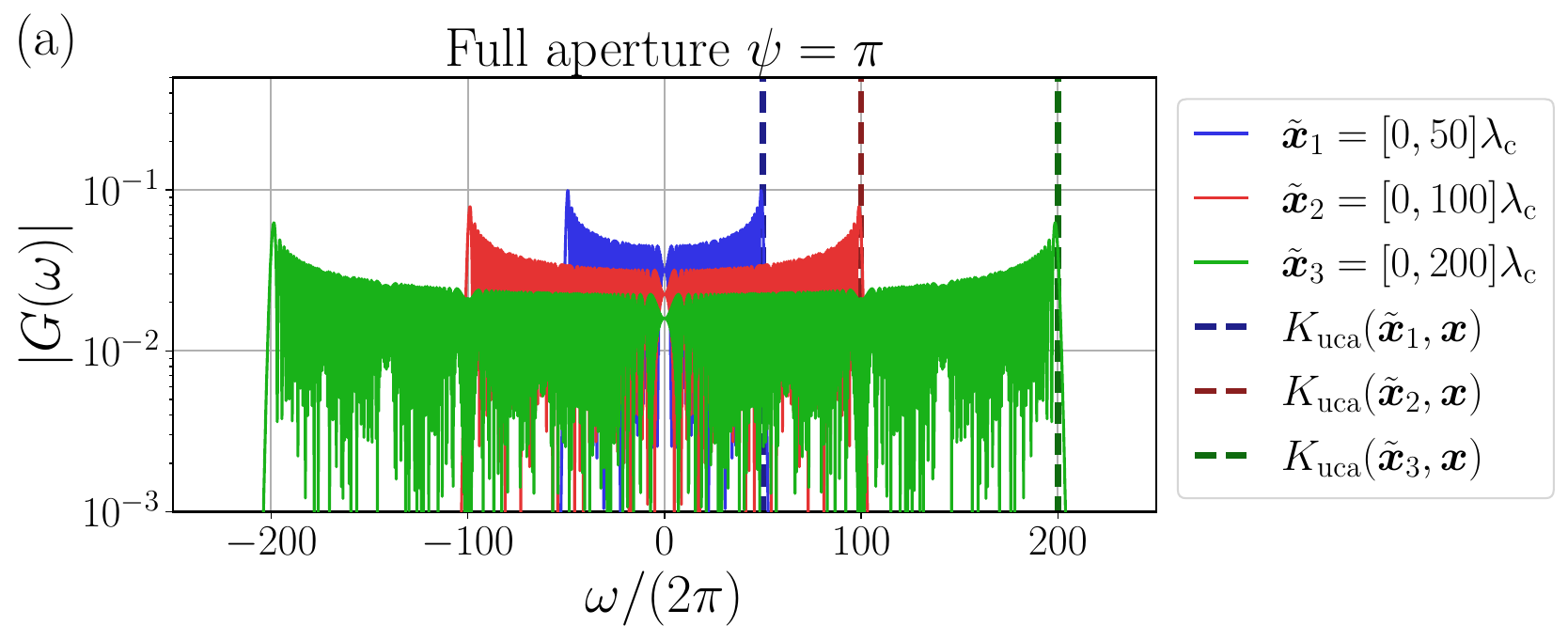}
    \includegraphics[width=\linewidth]{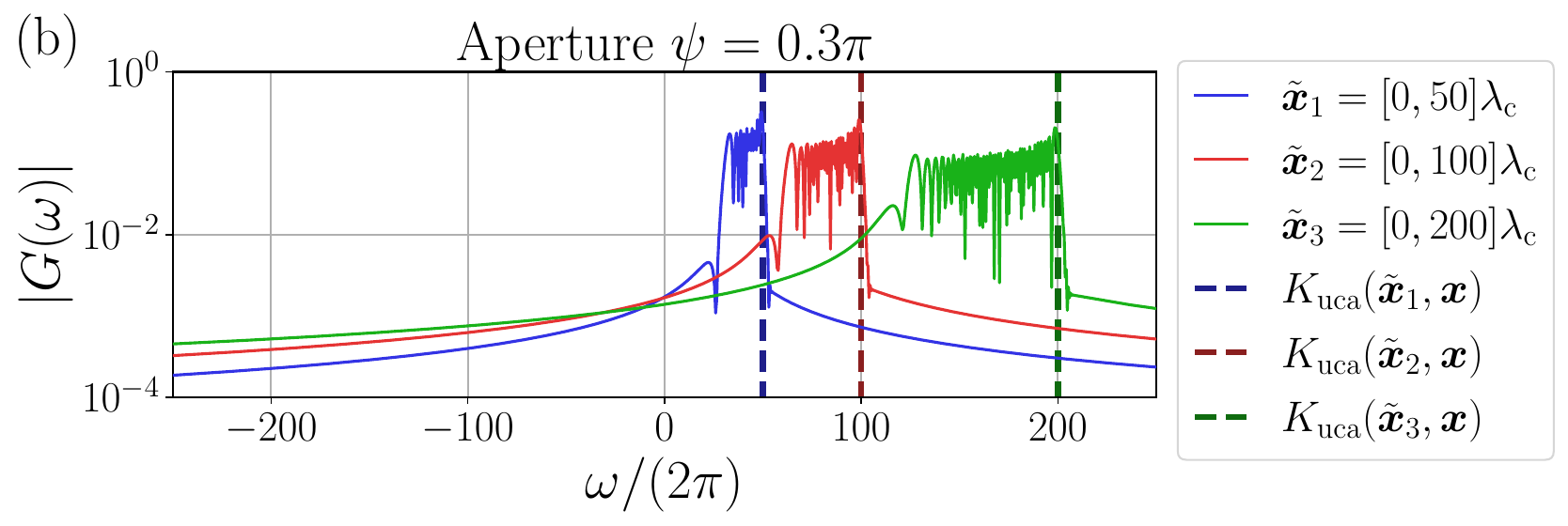}
    \includegraphics[width=\linewidth]{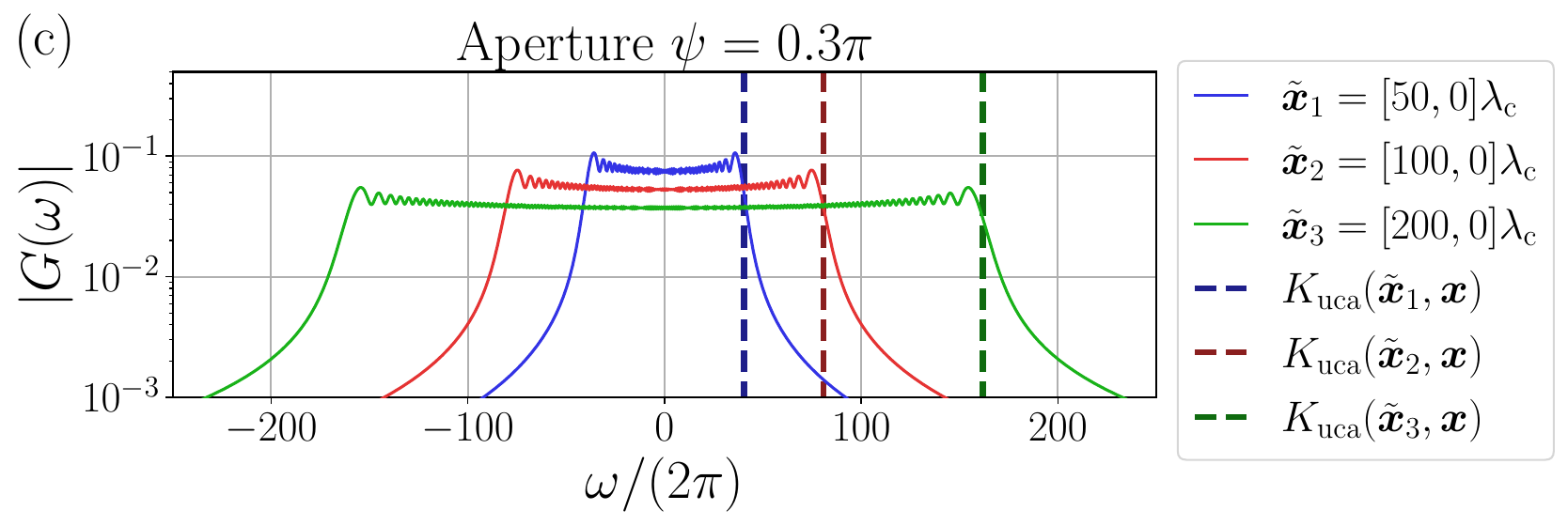}
    \caption{Spectrum $\matchspectrum(\paramfreq\seppar\testedloc, \sourceloc = \bs 0)$ for a \glspl{uca} with $\ucaradius=10,000\sourcewavelen$.
    (a) with $\aperture = \pi$.
    (b) with $\aperture = 0.3\pi$ and vertical position differences. 
    (c) with $\aperture = 0.3\pi$ and horizontal position differences.
    }
    \label{fig:uca-spectrum}
\end{figure}

Circular arrays constitute a natural reference for theoretical studies on \glspl{xla}~\cite{hafner_calibration_2019, vandendorpe_positioning_2025}, due to their inherent isotropic geometry providing a 360-degree coverage. 
As such, this topology can be considered as the canonical example of the ``perfectly" distributed \gls{xla}.
This simple geometry enables the derivation of closed-form expressions, exploiting the Fourier-Bessel series for its associated radiation patterns and ambiguity function, notably demonstrating a resolution with an order of magnitude comparable to the carrier wavelength~\cite{vandendorpe_positioning_2025, monnoyer_chirp_2025}.
While such results are associated with complete circles of continuous antenna elements, we aim here to exploit our framework to understand the aliasing structure that appears in the discrete-space \gls{af} of \glspl{uca}. 

In all generality, a \gls{uca} is described by an antenna domain $\antdomain$ that draws an arc on the zero-centered circle of radius $\ucaradius$ surrounding the operating domain $\sourcedomain$, as shown in Figure~\ref{fig:uca-scenario}.
The parametric path variable $\paramvar$ denotes here an angle such that 
\begin{equation}
    \paramtf(\paramvar)
    = \ucaradius
    \left[\begin{array}{c}
         \cos (\paramvar)  \\
         \sin (\paramvar)
    \end{array}\right],
\end{equation}
and we set the domain $\paramdomain = [-\aperture, \aperture]$ with $0\leq \aperture \leq \pi$ to enable us to consider either the full circle with $\aperture = \pi$ or arcs with $\aperture < \pi$.
The antenna spacing $\paramspacing$ corresponds, in the context of this section, to an \emph{angular} spacing, expressed in radians.

\begin{figure*}[!tb]
    \centering
    \includegraphics[width=\linewidth]{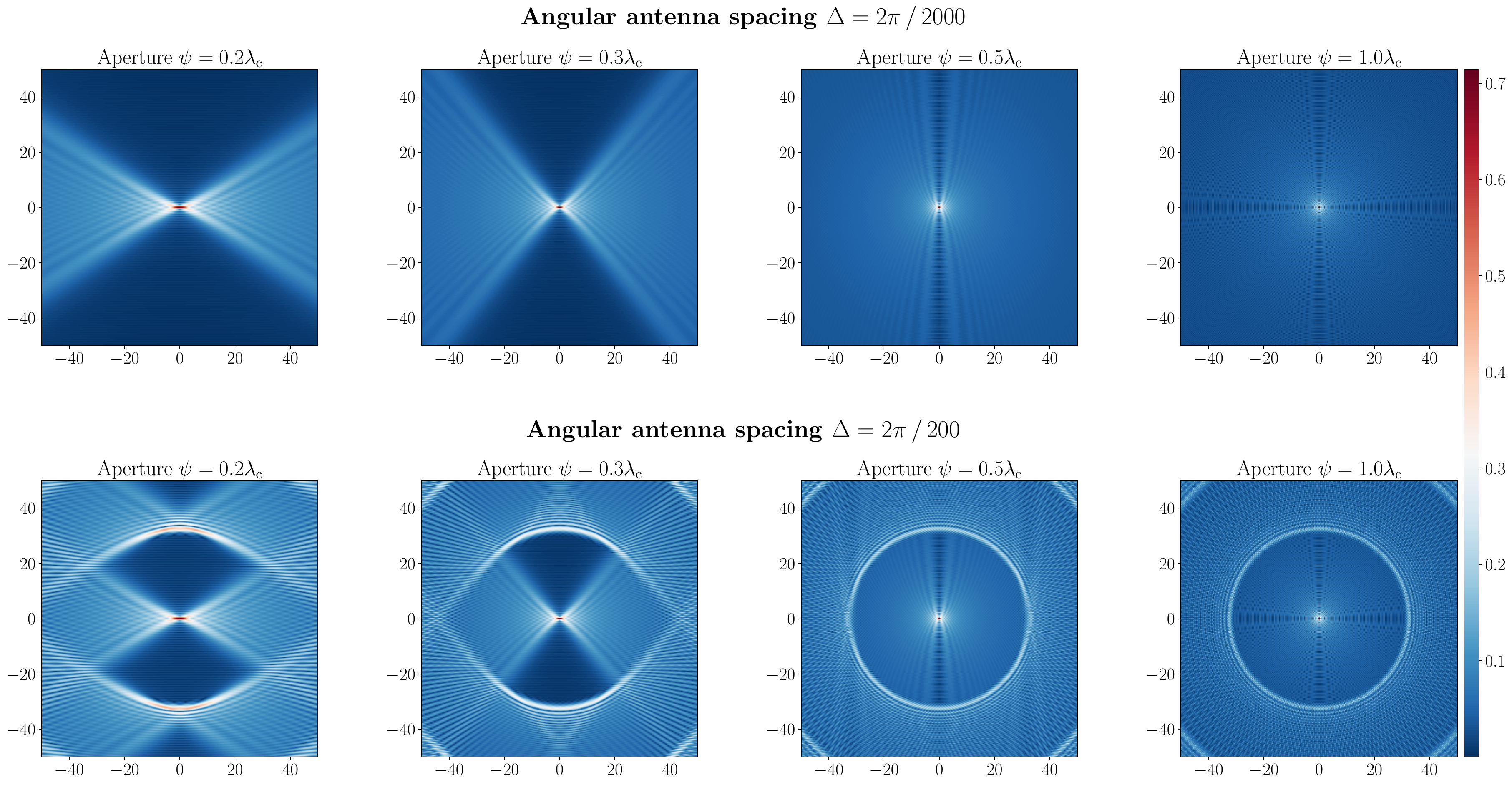}
    \caption{Discrete-space \glspl{af}, $|\sambifun(\testedloc, \sourceloc=\mathbf 0)|$, for a \gls{uca} of radius $\ucaradius = 10000 \sourcewavelen$ and multiple apertures $\aperture$. 
    (top) uniform angular spacing $\paramspacing = 2\pi / 2000$.
    (bottom) $\paramspacing = 2\pi/200$).
    }
    \label{fig:CA-AF}
\end{figure*}

To develop further the expression of the phase \eqref{eq:chirp-phase}, we restrict the focus of this section to an infinitely large array with $\ucaradius \rightarrow \infty$ surrounding the operating domain $\sourcedomain$.
Practically speaking, the resulting study provides good representations of the aliasing structure one observes for a non-infinite radius satisfying 
\begin{equation}
    \ucaradius \gg \max\{ \|\sourceloc\|: \sourceloc \in \sourcedomain \}.
\end{equation}
Under the infinite-radius condition, the matched signal's amplitude, given in \eqref{eq:chirp-amplitude}, is asymptotically constant, causing $\matchspectrum_\matchamp(\paramfreq)$ to be a Dirac delta.
As a result, $\matchspectrum(\paramfreq \seppar \testedloc, \sourceloc)$ and $\matchspectrum_\matchphase(\paramfreq)$ are exactly equals, up to a constant coefficient.

\medskip

The phase of this signal, given in \eqref{eq:chirp-phase}, remains non-linear across the infinite circle $\antdomain$, hence preserving the \gls{nf} nature of this scenario.
However, in the reciprocal point of view, each specific antenna's location $\antloc = \paramtf(\paramvar)$ is seen in the \gls{ff} of the domain $\sourcedomain$. 
Said differently, the domain $\sourcedomain$ is infinitely small compared to the distance separating it from $\antloc$.
Consequently, this condition validates the modeling in~\cite{vandendorpe_positioning_2025}, implying that for all $\antloc\in\antdomain$ and all $\testedloc, \sourceloc\in\sourcedomain$, the matched signal phase reads
\begin{equation}
    \label{eq:phase-uca-infty}
    \matchphase(\paramvar) 
    = \sourcek\diffradius \cos(\paramvar -\diffangle),
\end{equation}
where we define
\begin{equation}
    \diffradius := \|\sourceloc - \testedloc\| \text{ and }
    \diffangle := \angle (\sourceloc - \testedloc),
\end{equation}
as visualized in Figure~\ref{fig:uca-scenario}.
The resulting band limit is straightforwardly obtained and formalized in Theorem~\ref{thm:bandlimit-uca}.

\begin{theorem}
    \label{thm:bandlimit-uca}
    The infinite-radius \gls{uca} yields a soft band limit (Definition~\ref{def:soft-bandlimit}) that is described by the closed-form expression
    \begin{equation}
        \label{eq:bandlimit-uca}
        \bandlim_\uca(\testedloc, \sourceloc)
        = 
        \sourcek \diffradius\: \visualaperture(\diffangle),
    \end{equation}
    where $\visualaperture(\diffangle)$ is the \emph{visual aperture} of the array at the angle $\diffangle$. We define it as
    \begin{equation}
        \label{eq:visual-aperture}
        \visualaperture(\diffangle) := \ts \max_{\paramvar \in[-\aperture, \aperture]} |\sin(\paramvar-\diffangle)|.
    \end{equation}
\end{theorem}
\begin{proof}
    Using \eqref{eq:phase-uca-infty}, the local spatial frequency is trivially expressed as 
    $
        \dot\matchphase(\paramvar)
        =
        - \sourcek \diffradius \sin(\paramvar-\diffangle).
        \label{eq:spectrumca}
    $ 
    Developing the definition \eqref{eq:chirp-bandlimit} with this expression directly provides \eqref{eq:bandlimit-uca}.
\end{proof}
Let us note that \eqref{eq:visual-aperture} possesses a closed-form expression that we provide here for the sake of clarity. 
If there is an integer $n$ such that 
\begin{equation}
    \label{eq:uca-fulleye-condition}
    |\diffangle + (2n+1)\tfrac{\pi}{2}| \in [-\aperture, \aperture],
\end{equation}
then the visual aperture takes its maximum value $\visualaperture(\diffangle) = 1$.
Otherwise, the constraints in \eqref{eq:visual-aperture} are active, implying that 
\begin{equation}
    \label{eq:visual-aperture-line-zone}
    \visualaperture(\diffangle) = \max\{|\sin(\aperture+\diffangle), |\sin(\aperture-\diffangle)|\}.
\end{equation}

Fig.~\ref{fig:uca-spectrum} shows how expression \eqref{eq:bandlimit-uca} of the band limit properly captures the spectrum $\matchspectrum(\paramfreq \seppar \testedloc, \sourceloc)$ given a \gls{ue} located in $\sourceloc=0$, and multiple tested locations $\testedloc$. 
The complete \gls{uca} with $\aperture = \pi$ exhibits an angular invariance, implying only symmetric spectra, as seen in Figure~\ref{fig:uca-spectrum}.
This invariance no longer holds for incomplete circles.
Therefore, it results in asymmetric spatial spectra (Figure~\ref{fig:uca-spectrum}b), except for the particular case of horizontal positional difference $\testedloc-\sourceloc$ (Figure~\ref{fig:uca-spectrum}c), assuming an array oriented as depicted in Figure~\ref{fig:uca-scenario}.

We can now study the resulting \gls{afr} of the \gls{uca}, given the angular spacing $\paramspacing$. 
It follows the interpretable expression:
\begin{align}
    \afr_\uca(\sourceloc) &:= \{\testedloc : \bandlim_\uca(\testedloc, \sourceloc) \leq 2\pi \paramspacing^{-1}\} \\
    &= 
    \Big\{(\diffradius, \diffangle) : \diffradius \leq \frac{\sourcewavelen}{\paramspacing \visualaperture(\diffangle)}\Big\}.
    \label{eq:afr-uca-infty}
\end{align}
Expression \eqref{eq:afr-uca-infty} informs us that the aliasing front appears around $\sourceloc$, in each direction $\diffangle$, at a distance $\frac{\sourcewavelen}{\paramspacing \visualaperture(\diffangle)}$.
Interestingly, as soon as the \gls{uca} draws a \emph{half-circle}, \ie when $\aperture \geq \frac{\pi}{2}$, condition~\eqref{eq:uca-fulleye-condition} is met, and hence the visual aperture provides its maximum value $\visualaperture(\diffangle) =1$ for all $\diffangle$.
As a result, the aliasing front draws a perfect circle around $\sourceloc$.

By contrast, when $\aperture < \frac{\pi}{2}$, then $\visualaperture(\diffangle) =1$ only for angles satisfying \eqref{eq:uca-fulleye-condition}.
These are angles such that the line passing through $\sourceloc$ that is normal to them crosses the array.
In other angles, the visual aperture follows \eqref{eq:visual-aperture-line-zone}, resulting in a locally linear aliasing front in the \gls{af}.

These effects can be observed in Figure~\ref{fig:CA-AF}, which shows discrete-space \glspl{af} computed for \glspl{uca} with multiple apertures $\sourcewavelen$ and either $\paramspacing = (2\pi) / 2000$ or $\paramspacing = (2\pi) / 200$ as angular antenna spacing.
The denser array acts as the aliasing-free reference.
We indeed observe a circular aliasing front of radius $\frac{\sourcewavelen}{\paramspacing} = 31.83 \sourcewavelen$ with alternating fronts and decaying zones repeating at every multiple of this radius, 
This pattern aligns with the analysis we presented in Figure~\ref{fig:Toy-AF-Zoom}.
When $\aperture < \frac{\pi}{2}$, the circle stretches as the visual aperture increases.
The portion of the front that fits the circle exactly corresponds to the angular aperture of the array (\ie an arc of $2\aperture$ radians). 
These two arcs---upper and lower--- are oriented perpendicularly to the array. 
Notably, the resulting eye-shaped region always includes the full-aperture disc, in accordance with Property~\ref{thm:inclusion-ii}.
That is, because any arc of aperture $2\aperture$ inherently represents a truncation of the whole circle.

%% file: sections/7-Conclusions.tex
\glsresetall

\section{Conclusions}
This paper has introduced a novel theoretical framework to address the challenging characterization of \gls{nf} grating lobes.
In the proposed approach, we have formally defined these lobes as aliasing artifacts affecting \glspl{af}.
Their structural origin has been further explained by leveraging a local spatial-frequency representation of the steering and matched signals. 
This enabled us to extend the conventional \gls{ff} interpretation of grating lobes while consistently explaining their observed behavior in the \gls{nf} regime.

Exploiting this understanding, this paper has provided a systematic methodology to derive \emph{\glspl{afr}}. 
The boundaries of those regions analytically define the geometry of the \gls{nf} grating lobes.
Studying them is the key to designing \glspl{asod}.
These domains guarantee no aliasing when restricting a given array to operate in this domain. 
Deriving methods to mathematically determine \glspl{asod} therefore constitutes a potential next step that directly follows the current work. 

This paper illustrated its practical value in deriving closed-form expressions of \glspl{afr} for both \glspl{ula} and \glspl{uca}.
In the context of \glspl{ula}, the present study has also detailed the connection between its \gls{nf} findings and established \gls{ff} knowledge.
This result further illustrated how this work can potentially bridge this gap in a unified and coherent framework.

Importantly, the proposed framework is currently restricted to explaining the geometrical \emph{presence} of aliasing effects--and hence grating lobes-- in the \glspl{af} of \gls{nf} arrays.
As a key feature of the \gls{nf} regime, these lobes are usually weaker in amplitude than the main lobe.
This allows \glspl{xla} to theoretically function beyond the \gls{afr}. 
Expanding the present analysis to include aliasing amplitudes is therefore critical to fully assess the attainable performance of \glspl{xla}.
Moreover, subsequent work extending our findings to include the resolution of both the main and grating lobes represents a key step toward a comprehensive characterization of the \gls{af} properties.

Future work directions also include exploiting the flexibility of our approach to determine the attainable \glspl{afr} for more sophisticated array topologies that meet the demands of practical applications.
This is expected to be achievable because the current methodology exploits a parametric representation theoretically capable of describing any antenna configuration.
However, applying this representation to highly complex topologies may require adapting the present framework to assess the practical significance of the grating lobes. 
This challenge emphasizes the importance of including the amplitude characterization of the aliasing artifacts in the analysis.

Finally, this work is expected to be adaptable to study the coupled space-time-frequency \glspl{af} arising when considering wideband systems and Doppler effects.
This adaptability paves the way for additional extensions, defining a general methodology to study the \gls{nf} grating lobes in wideband non-static scenarios.  

%% file: sections/8-Misc.tex
\section*{Acknowledgments}
L. Defraigne and B. Sambon thank the FRIA and the FRS-FNRS for their financial support

\section{Code Availabiility}
All relevant code supporting the document is available here:\\ 
{\small \url{https://forge.uclouvain.be/GillesMonnoyer/NPJWT-Aliasing-NF-AF}.}\\
For any additional request, please refer to G.M., gilles.monnoyer@uclouvain.be.

\section*{Author contributions}
G. M. and L. V. came to the idea of investigating the impact of array discretization on the ambiguity function through the approach of spatial frequencies, and the formal link between \gls{af} and spectral content. 
G. M. introduced the \gls{afr} and \gls{asod} notions. 
G. M. achieved the theoretical results for the \gls{ula}. 
L. V. derived the results for the \gls{uca}. 
All numerical results were generated by G. M., as well as the first version of the manuscript. 
J.L, L.D., and B.S. highly contributed to the ordering of ideas, the identification of the best ways to present them, and to the improvement of the text.

\section*{Competing Interests}
There are no competing interests to declare.

%% file: sections/9-Appendix.tex
\appendix

\subsection{Proof of Theorem \ref{thm:lam2}}
    The first-order derivative of \eqref{eq:chirp-phase} is given by
    \begin{equation}
        \dot\matchphase(\paramvar) = \sourcek
        \bigg(
        \frac{\paramtf(\paramvar) - \testedloc}{\|\paramtf(\paramvar) - \testedloc\|}
        -
        \frac{\paramtf(\paramvar) - \sourceloc}{\|\paramtf(\paramvar) - \sourceloc\|}
        \bigg)
        \,
        \dot\paramtf(\paramvar).
    \end{equation}
    Using the Cauchy-Schwarz inequality, we conclude that
    \begin{equation}
        \label{eq:demo-lam2-cauchy}
        |\dot\matchphase(\paramvar)|
        \leq
        \sourcek
        \bigg\|\frac{\paramtf(\paramvar) - \testedloc}{\|\paramtf(\paramvar) - \testedloc\|}
        -
        \frac{\paramtf(\paramvar) - \sourceloc}{\|\paramtf(\paramvar) - \sourceloc\|}
        \bigg\|
        \,
        \|\dot\paramtf(\paramvar)\|.
    \end{equation}
    The first factor in the right-side product in \eqref{eq:demo-lam2-cauchy} is the norm of a difference of two unit-norm vectors, which is upper-bounded by $2$.
    As a result, $\bandlim(\testedloc, \sourceloc) \leq 2\sourcek \max_{\paramvar \in \paramdomain}\|\dot\paramtf(\paramvar)\|$ for any pair $\testedloc$, $\sourceloc$.
    Since $\sourcek = \frac{2\pi}{\sourcewavelen}$, the condition \eqref{eq:afr} is always met when 
    \begin{equation}
        \paramspacing \max_{\paramvar \in \paramdomain}\|\dot\paramtf(\paramvar)\| \leq \frac{\sourcewavelen}{2}.
    \end{equation}
    Finally, from the mean-value theorem, for any pair $\paramvar_1, \paramvar_2 \in \paramdomain$, we have
    \begin{equation}
        \|\paramtf(\paramvar_2) - \paramtf(\paramvar_1)\| \leq |\paramvar_2 - \paramvar_1| \max_{\paramvar \in \paramdomain}\|\dot\paramtf(\paramvar)\|.
    \end{equation}
    Therefore, \eqref{eq:thm-lam2-condition} validates the condition in the definition of the \gls{afr} for all $\testedloc$, $\sourceloc \in \bb R^\ndim$, from which we directly conclude that any set $\sourcedomain\subseteq \bb R^\ndim$ is an \gls{asod}.
    This concludes this proof.

\subsection{Proof of Theorem~\ref{thm:bandlimit-ula}}
\label{app:ula}
Particularizing Definition~\ref{def:soft-bandlimit} to the infinite-length \gls{ula} described in Section~\ref{sec:ula} yields
\begin{align}
    \label{eq:bandlimit-ula-base}
    \bandlim_\infty(\testedloc, \sourceloc) 
    &:= 
    \max_{\paramvar \in \bb Z}
    |\dot\matchphase(\paramvar)|.
\end{align}
Developing the expression of $\dot\matchphase(\paramvar)$ with the above locations leads to
\begin{equation}
    \label{eq:freq-ula}
    \dot\matchphase(\paramvar) = \sourcek \left[\frac{\paramvar - \xloc}{((\paramvar-\xloc)^2 + \yloc^2)^\frac{1}{2}} - \frac{\paramvar - \xtest}{((\paramvar - \xtest)^2 + \ytest^2)^\frac{1}{2}}\right],
\end{equation}
and its derivative is
\begin{equation}
    \label{eq:freq-der-ula}
    \ddot\matchphase(\paramvar) = \sourcek \left[\frac{\yloc^2}{((\paramvar - \xloc)^2 + \yloc^2)^\frac{3}{2}} - \frac{\ytest^2}{((\paramvar - \xtest)^2 + \ytest^2)^\frac{3}{2}}\right].
\end{equation}

Since we assume $\yloc > 0$ and $\ytest > 0$ (excluding the source to be located inside and behind the antenna array), the function $\dot\matchphase(\paramvar)$ in \eqref{eq:freq-ula} is continuous as both denominators are always non-zero.
A direct application of the L'Hospital's rule shows that this function tends toward 0 in both $-\infty$ and $+\infty$.
Therefore, the maximizer of \eqref{eq:bandlimit-ula-base} is necessarily given by (one of) the roots of $\ddot\matchphase(\paramvar)$.
Denoting this maximizer by $\prootula$, we have
\begin{equation}
    \label{eq:bandlimit-ula-simple}
    \bandlim_\infty(\testedloc, \sourceloc) = |\dot\matchphase(\prootula)|.
\end{equation}

Since $\prootula$ is one root of $\ddot\matchphase(\paramvar)$, it must satisfy
\begin{equation}
    \label{eq:ula-root-cond}
    \ytest^\frac{4}{3} ((\prootula - \xloc)^2 + \yloc^2) = \yloc^\frac{4}{3}((\prootula - \xtest)^2 + \ytest^2).
\end{equation}
Before solving \eqref{eq:ula-root-cond}, let us simplify its expression by introducing the normalization
\begin{equation}
    \label{eq:normalized-root-def}
    \prootulan := \frac{\prootula - \xloc}{\yloc},
\end{equation}
and by exploiting the notations introduced in \eqref{eq:ula-bandlim-ratio}. 
After a few algebraic manipulations, we rewrite \eqref{eq:ula-root-cond} as
\begin{equation}
    \label{eq:normalized-ula-root-cond}
    (\prootulan - \xratio)^2 + \yratio^3 = \yratio^2 (\prootulan^2 +1).
\end{equation}

The above defines a second-order polynomial in $\prootulan$, whose roots are 
\begin{equation}
    \label{eq:normalized-root}
    \prootulan = \frac{
    q \yratio \sqrt{(\yratio^2-1)(\yratio-1) + \xratio^2 }
    - \xratio
    }{
    \yratio^2-1
    }.
\end{equation}
for $q\in\{-1, 1\}$.
The above square root's content is always positive since $\yratio>0$ under our conditions.

Since \eqref{eq:normalized-root-def} amounts to write $\prootula = \yloc \prootulan + \xloc$, we must now evaluate $\dot\matchphase(\yloc \prootulan + \xloc)$ and determine the sign of $q$ that yields maximality.
To this end, let us first similarly apply the notations introduced in \eqref{eq:ula-bandlim-ratio} and \eqref{eq:normalized-root-def} in order to repress \eqref{eq:freq-ula}.
With a few algebraic manipulations, we obtain
\begin{equation}
    \label{eq:normalized-freq-ula}
    \dot\matchphase(\yloc \prootulan + \xloc) = \sourcek \left[\frac{\prootulan}{(\prootulan^2 + 1)^\frac{1}{2}} - \frac{\prootulan - \xratio}{((\prootulan - \xratio)^2 + \yratio^3)^\frac{1}{2}}\right],
\end{equation}
Now substituting the identity \eqref{eq:normalized-ula-root-cond} in \eqref{eq:normalized-freq-ula} enables the simplification
\begin{equation}
    \dot\matchphase(\yloc \prootulan + \xloc) = \sourcek \frac{(\yratio - 1) \prootulan + \xratio}{\yratio(\prootulan^2 + 1)^\frac{1}{2}}.
\end{equation}

Exploiting \eqref{eq:bandlimit-ula-simple}, we conclude that there is at least one sign $q$ such that
\begin{align}
    \label{eq:bandlimit-ula-endproof}
    \bandlim_\infty(\testedloc, \sourceloc) = |\dot\matchphase(\yloc \prootulan + \xloc)|
    &= 
    \sourcek
    \bigg|
    \frac{(\yratio - 1) \prootulan + \xratio}{\yratio(\prootulan^2 + 1)^\frac{1}{2}}
    \bigg| \\
    &=
    \sourcek
    \frac{|(\yratio - 1) \prootulan + \xratio|}{\yratio(\prootulan^2 + 1)^\frac{1}{2}},
\end{align}
which is exactly \eqref{eq:thm-bandlimit-ula}.

The next step of this proof is to find the sign $q$ that maximizes $|\dot\matchphase(\yloc \prootulan + \xloc)|$. 
To this end, we use \eqref{eq:normalized-root} to develop \eqref{eq:bandlimit-ula-endproof}.
Proceeding separately for the numerator and the denominator, we have
\begin{equation}
    (\yratio - 1) \prootulan + \xratio
    =
    \frac{\yratio}{\yratio + 1} (q\sqrt{(\yratio^2-1)(\yratio-1) + \xratio^2 } + \xratio),
\end{equation}
and  
\begin{multline}
    \prootulan^2 +1
    =
    \frac{1}{\yratio^2 -1} 
    \big(
    (\yratio^2-1)(\yratio^3-1) + \xratio^2(\yratio^2+1) 
    \\
    - 2q \yratio \xratio \sqrt{(\yratio^2-1)(\yratio-1) + \xratio^2 }
    \big).
\end{multline}
Reminding that $\yratio>0$, we conclude that the numerator and denominator of \eqref{eq:bandlimit-ula-endproof} are respectively maximized and minimized, in absolute value, when $q = \sign(\xratio)$, thereby guaranteeing that this sign maximizes $|\dot\matchphase(\yloc \prootulan + \xloc)|$. 
Injecting this sign in \eqref{eq:normalized-root} and applying the change of variable $\xratio = (\yratio^2-1) \xyratio$ yields \eqref{eq:thm-bandlimit-ula-root} and concludes this proof.

\subsection{Proofs of Corollaries~\ref{cor:bandlimit-ula-horizontal} and~\ref{cor:bandlimit-ula-vertical}}
\label{app:ula-cor}
\begin{proof}[proof of corollary \ref{cor:bandlimit-ula-horizontal}]
    Starting from \eqref{eq:normalized-ula-root-cond} with $\yratio = 1$, we have
    \begin{equation}
        (\prootulan - \xratio)^2 + 1 = \prootulan^2 +1,
    \end{equation}
    which simplifies to $\prootulan = \xratio/2$ (and thus the maximizer is $\prootula = \yloc \xratio / 2 + \xloc$).
    The expression \eqref{eq:thm-bandlimit-ula} therefore simplifies into \eqref{eq:thm-bandlimit-ula-u1}, concluding this proof.
\end{proof}
\begin{proof}[proof of corollary \ref{cor:bandlimit-ula-vertical}]
    When $\xratio = \xyratio = 0$, the expression \eqref{eq:thm-bandlimit-ula-root} is reduced to
    \begin{equation}
        \prootulan = \sign(\xyratio) \yratio / \sqrt{\yratio+1}.
    \end{equation}
    The expression \eqref{eq:thm-bandlimit-ula} is in turns reduced to
    \begin{equation}
        \bandlim_\infty(\testedloc, \sourceloc)
        =
        \tfrac{\yratio}{\sqrt{\yratio+1}} \Big/ \sqrt{1 + \tfrac{\yratio^2}{\yratio+1}},
    \end{equation}
    which simplifies into \eqref{eq:thm-bandlimit-ula-v0}, concluding this proof.
\end{proof}

\subsection{Proofs of ULA's Aliasing-free Regions Properties}
\label{app:ula-afr-prop}
\begin{proof}[Proof of Property~\ref{prop:ula-inv}]
    Let us consider the pair of $\testedloc_1 = [\xtest_1, \ytest_1]^\top$, $\sourceloc_1 = [0, 1]^\top$, from which we construct another as $\testedloc_2 = [\yloc_2\xtest_1 + \xloc_2, \yloc_2\ytest_1]^\top$, $\sourceloc_2 = [\xloc_2, \yloc_2]^\top$.
    This construction always yields
    \begin{equation}
        \label{eq:bandlin-ula-translation}
        \bandlim_\infty(\testedloc_1, \sourceloc_1) 
        =
        \bandlim_\infty(\testedloc_2, \sourceloc_2)
    \end{equation}
    because both pairs give rise to the same values for $\yratio$ and $\xratio$.
    Therefore for all $\testedloc_1 \in \afr_\infty(\sourceloc_1)$, the second pair meets $\testedloc_2 \in \afr_\infty(\sourceloc_2)$, and inversely.
    By construction the equality $\afr_\infty(\sourceloc_1) = \ulaeye$ holds. Since we can also write 
    \begin{equation}
        \testedloc_2 = \yloc_2 \cdot \testedloc_1 + \left[\begin{array}{c}
             \xloc_2 \\
             0 
        \end{array}\right],
    \end{equation}
    it concludes the proof.
\end{proof}

\begin{proof}[Proof of Property~\ref{prop:ula-symmetry}]
    Since $\ulaeye$ is obtained from Theorem~\ref{thm:bandlimit-ula} with $\xloc=0$ and $\yloc=1$, inverting the sign of $\xtest$ inverts the sign of $\xratio$ and hence also inverts the sign of $\prootulan$, while maintaining their absolute value.
    Therefore, both the numerator and the denominator in \eqref{eq:thm-bandlimit-ula} are unchanged in absolute value.
    It results in  
    \begin{equation}
        \bandlim_\infty([-\xtest, \ytest]^\top, [0, 1]^\top) = \bandlim_\infty([\xtest, \ytest]^\top, [0, 1]^\top),
    \end{equation}
    for all $\testedloc$.
    Therefore, given $\sourceloc=[0,1]^\top$ if $[\xtest, \ytest]^\top$ satisfies \eqref{eq:afr-ula-infty}, then $[-\xtest, \ytest]^\top$ satisfies it too.
\end{proof}

\begin{proof}[Proof of Property~\ref{prop:ula-eye-aperture}]
    We begin this proof with the expression \eqref{eq:eyewidth} of the width $\eyewidth$.
    Starting from \eqref{eq:thm-bandlimit-ula-u1}, we have that the boundary of $\afr_\infty(\sourceloc)$ is met in that direction when $\spacingratio|\xratio| = \sqrt{\xratio^2/4 +1}$, which is trivially met when $\xratio = 2 / \sqrt{4\spacingratio^2 -1}$.
    Note that in the context of the eye $\ulaeye$, we have $\xratio = \xtest$, thereby proving \eqref{eq:eyewidth}.

    We now prove the vertical apertures using \eqref{eq:thm-bandlimit-ula-v0}.
    Similarly, the boundary of $\afr_\infty(\sourceloc)$ is found when $\spacingratio|\yratio -1| = \sqrt{1 + \yratio + \yratio^2}$.
    Squaring both sides yields
    \begin{equation}
        \label{eq:ula-v0-polynomial}
        \paramspacing^2(\yratio^2-1)^2 = 1 + \yratio + \yratio^2,
    \end{equation}
    which finds a solution in
    \begin{equation}
        \yratio = \frac{2\spacingratio^2 +1 +q\sqrt{12 \spacingratio^2 -3}}{2(\spacingratio^2-1)},
    \end{equation}
    for $q\in \{-1, 1\}$. 
    When $\spacingratio^2 = 1$, then \eqref{eq:ula-v0-polynomial} finds 0 as its only solution. 
    When $\frac{3}{4} \leq \spacingratio^2 < 1$, then \eqref{eq:ula-v0-polynomial} finds only negative solutions which must be excluded because Theorem~\ref{thm:bandlimit-ula} is only valid for positive $\yratio$ (upper half plane only). 
    When $\spacingratio^2 < \frac{3}{4}$, the polynomial equation \eqref{eq:ula-v0-polynomial} finds no real solution. 
    In all those cases, the eye's aperture vertically degenerates in 0 (on the bottom) and $+\infty$ (on top). 
    Otherwise, the solution with $q=1$ gives the top border and the one with $q=-1$, the lower border. 
    Finally, noticing that, in the context of the eye $\ulaeye$, we have $\yratio = \ytest^\frac{2}{3}$ leads to \eqref{eq:eyetop} and \eqref{eq:eyebot}, which concludes this proof.
\end{proof}

\rev{
\subsection{Proof of Theorem~\ref{thm:bandlinit-ula-L}}
\label{proof:thm-ula-L}
This proof follows that of Theorem~\ref{thm:bandlimit-ula} and uses the same notations.  
    We aim to solve
        \begin{align}
        \label{eq:bandlimit-ula-base-2}
        \bandlim_\lenula(\testedloc, \sourceloc) 
        &:= 
        \max_{\paramvar \in [-\tfrac{\lenula}{2}, \tfrac{\lenula}{2}]}
        |\dot\matchphase(\paramvar)|.
    \end{align}
    When the unconstrained maximizer, $\yloc\prootulan + \xloc$, is in $[-\tfrac{\lenula}{2}, \tfrac{\lenula}{2}]$, it is also the maximizer of \eqref{eq:bandlimit-ula-base-2}, thereby demonstrating the first case in \eqref{eq:thm2-bandlim-ula}.    
    
    When the constraint excludes $\yloc\prootulan + \xloc$, the solution can either be given by the only other local extremum of $\dot\matchphase(\paramvar)$ or by evaluating it at one edge of the constraint. 
    That is, since we showed in Theorem~\ref{thm:bandlimit-ula}'s proof that $\dot\matchphase(\paramvar)$ is continuous, tends to $0$ when $\paramvar \mapsto \pm \infty$, and possesses at most two local extrema, in $\yloc\prootulan + \xloc$ and $\yloc\prootulansec + \xloc$. 
    This situation can be separated into two cases, respectively providing the second and third regimes in \eqref{eq:thm2-bandlim-ula}. 
    
    \paragraph*{case $\yloc\prootulansec + \xloc \in [-\tfrac{\lenula}{2}, \tfrac{\lenula}{2}]$}
    As $\dot\matchphase(\paramvar)$ tends to $0$ when $\paramvar \mapsto \pm \infty$, this case necessarily implies that $|\dot\matchphase(\yloc\prootulansec + \xloc)| > |\dot\matchphase(-\sign(\xyratio)\tfrac{\lenula}{2})|$. 
    The two remaining candidates to maximize \eqref{eq:bandlimit-ula-base-2} are therefore $\paramvar=\yloc\prootulansec + \xloc$ and $\paramvar=\sign(\xyratio)\tfrac{\lenula}{2}$.
    Evaluating $|\dot\matchphase(\cdot)|$ in these values provides, after a few algebraic manipulations, the expressions of $\bandlimplus(\testedloc, \sourceloc)$ and $\bandlimsec(\testedloc, \sourceloc)$, respectively given in \eqref{eq:bandlimpm} and \eqref{eq:bandlimsec}
    
    \paragraph*{case ``\emph{otherwise}"}
    In that case, the only candidates to maximize \eqref{eq:bandlimit-ula-base-2} are the two edges of the array, \ie $\paramvar = \pm \tfrac{\lenula}{2}$, which directly yields $\bandlim_\lenula (\testedloc, \sourceloc) = \max\{\bandlimplus(\testedloc, \sourceloc), \bandlimminus(\testedloc, \sourceloc)\}$ and concludes this proof.
}

%% file: nourl.bib
@IEEEtranBSTCTL{MyBSTcontrol,
    CTLuse_url = "no",
}


%% file: references-NF-SL.bib
@article{liu_near_field_2023,
	title = {Near-{Field} {Communications}: {A} {Tutorial} {Review}},
	volume = {4},
	issn = {2644-125X},
	shorttitle = {Near-{Field} {Communications}},
	url = {https://ieeexplore.ieee.org/document/10220205},
	doi = {10.1109/OJCOMS.2023.3305583},
	urldate = {2025-04-11},
	journal = {IEEE Open Journal of the Communications Society},
	author = {Liu, Yuanwei and Wang, Zhaolin and Xu, Jiaqi and Ouyang, Chongjun and Mu, Xidong and Schober, Robert},
	year = {2023},
	pages = {1999--2049},
}

@article{chen_6g_2024,
  journal={IEEE Wireless Communications}, 
  title={{6G} {Localization} and {Sensing} in the {Near} {Field}: {Features}, {Opportunities}, and {Challenges}}, 
  author={Chen, Hui and Keskin, Musa Furkan and Sakhnini, Adham and Decarli, Nicolò and Pollin, Sofie and Dardari, Davide and Wymeersch, Henk},
  year={2024},
  volume={31},
  number={4},
  pages={260-267},
  doi={10.1109/MWC.011.2300359}}

@article{demir_foundations_2021,
	title = {Foundations of {User}-{Centric} {Cell}-{Free} {Massive} {MIMO}},
	volume = {14},
	issn = {1932-8346, 1932-8354},
	url = {https://www.nowpublishers.com/article/Details/SIG-109},
	doi = {10.1561/2000000109},
	language = {English},
	number = {3-4},
	urldate = {2025-04-11},
	journal = {SIG},
	author = {Demir, Ozlem Tugfe and Bjornson, Emil and Sanguinetti, Luca},
	month = jan,
	year = {2021},
	pages = {162--472},
}

@article{wu_multiple_2023,
	title = {Multiple {Access} for {Near}-{Field} {Communications}: {SDMA} or {LDMA}?},
	volume = {41},
	issn = {1558-0008},
	shorttitle = {Multiple {Access} for {Near}-{Field} {Communications}},
	url = {https://ieeexplore.ieee.org/document/10123941},
	doi = {10.1109/JSAC.2023.3275616},
	number = {6},
	urldate = {2025-04-11},
	journal = {IEEE Journal on Selected Areas in Communications},
	author = {Wu, Zidong and Dai, Linglong},
	month = jun,
	year = {2023},
	pages = {1918--1935},
}

@inproceedings{swindlehurst_passive_1988,
	title = {Passive direction-of-arrival and range estimation for near-field sources},
	url = {https://ieeexplore.ieee.org/document/206176},
	doi = {10.1109/SPECT.1988.206176},
	urldate = {2025-04-11},
	booktitle = {Fourth {Annual} {ASSP} {Workshop} on {Spectrum} {Estimation} and {Modeling}},
	author = {Swindlehurst, A.L. and Kailath, T.},
	month = aug,
	year = {1988},
	pages = {123--128},
}

@inproceedings{qiu_doa_2018,
	title = {{DOA} {Estimation} of {Near}-field {Passive} {Sources} with {Acoustic} {Array} {Based} on {Fractional} {Fourier} {Transform}},
	url = {https://ieeexplore.ieee.org/document/8604699},
	doi = {10.1109/OCEANS.2018.8604699},
	urldate = {2025-04-11},
	booktitle = {{OCEANS} 2018 {MTS}/{IEEE} {Charleston}},
	author = {Qiu, Wei and Wang, Wenke and Xiao, Wenbin},
	month = oct,
	year = {2018},
	pages = {1--4},
}

@inproceedings{jian_fractional_2024,
	title = {Fractional {Fourier} {Transformation} {Based} {XL}-{MIMO} {Near}-{Field} {Channel} {Analysis}},
	url = {https://ieeexplore.ieee.org/document/10694136},
	doi = {10.1109/SPAWC60668.2024.10694136},
	urldate = {2025-04-11},
	booktitle = {2024 {IEEE} 25th {International} {Workshop} on {Signal} {Processing} {Advances} in {Wireless} {Communications} ({SPAWC})},
	author = {Jian, Mengnan and Tang, Anzheng and Chen, Yijian and Zhao, Yajun},
	month = sep,
	year = {2024},
	pages = {221--225},
}

@inproceedings{chassande_stationary_1998,
	title = {On the stationary phase approximation of chirp spectra},
	url = {https://ieeexplore.ieee.org/document/721375},
	doi = {10.1109/TFSA.1998.721375},
	urldate = {2025-04-11},
	booktitle = {Proceedings of the {IEEE}-{SP} {International} {Symposium} on {Time}-{Frequency} and {Time}-{Scale} {Analysis} ({Cat}. {No}.{98TH8380})},
	author = {Chassande-Mottin, E. and Flandrin, P.},
	month = oct,
	year = {1998},
	pages = {117--120},
}

@inproceedings{vandendorpe_positioning_2025,
	title = {Positioning and transmission in cell-free networks: ambiguity function, and {MRC}/{MRT} array gains},
	shorttitle = {Positioning and transmission in cell-free networks},
	url = {https://ieeexplore.ieee.org/document/10889929},
	doi = {10.1109/ICASSP49660.2025.10889929},
	urldate = {2025-04-11},
	booktitle = {{ICASSP} 2025 - 2025 {IEEE} {International} {Conference} on {Acoustics}, {Speech} and {Signal} {Processing} ({ICASSP})},
	author = {Vandendorpe, Luc and Defraigne, Laurence and Thiran, Guillaume and Pairon, Thomas and Craeye, Christophe},
	month = apr,
	year = {2025},
	pages = {1--5},
}

@article{abhayapala_spatial_2002,
	title = {Spatial {Aliasing} for {Nearfield} {Sensor} {Arrays}},
	journal = {Electronics Letters},
	author = {Abhayapala, Thushara and Kennedy, Rodney and Williamson, Robert},
	month = aug,
	year = {2002},
}

@article{gui_generalized_2022,
	title = {Generalized {Ambiguity} {Function} for {FDA} {Radar} {Joint} {Range}, {Angle} and {Doppler} {Resolution} {Evaluation}},
	volume = {19},
	issn = {1558-0571},
	url = {https://ieeexplore.ieee.org/abstract/document/9309182},
	doi = {10.1109/LGRS.2020.3044351},
	urldate = {2025-06-26},
	journal = {IEEE Geoscience and Remote Sensing Letters},
	author = {Gui, Ronghua and Huang, Bang and Wang, Wen-Qin and Sun, Yan},
	year = {2022},
	pages = {1--5},
}

@inproceedings{li_near_field_2023,
	title = {Near-{Field} {Beam} {Focusing} {Pattern} and {Grating} {Lobe} {Characterization} for {Modular} {XL}-{Array}},
	url = {https://ieeexplore.ieee.org/abstract/document/10436773},
	doi = {10.1109/GLOBECOM54140.2023.10436773},
	urldate = {2025-07-05},
	booktitle = {{GLOBECOM} 2023 - 2023 {IEEE} {Global} {Communications} {Conference}},
	author = {Li, Xinrui and Dong, Zhenjun and Zeng, Yong and Jin, Shi and Zhang, Rui},
	month = dec,
	year = {2023},
	note = {ISSN: 2576-6813},
	pages = {4068--4073},
}

@article{wang_extremely_2024,
	title = {Extremely {Large}-{Scale} {MIMO}: {Fundamentals}, {Challenges}, {Solutions}, and {Future} {Directions}},
	volume = {31},
	issn = {1558-0687},
	shorttitle = {Extremely {Large}-{Scale} {MIMO}},
	url = {https://ieeexplore.ieee.org/abstract/document/10098681},
	doi = {10.1109/MWC.132.2200443},
	number = {3},
	urldate = {2025-08-15},
	journal = {IEEE Wireless Communications},
	author = {Wang, Zhe and Zhang, Jiayi and Du, Hongyang and Sha, Wei E. I. and Ai, Bo and Niyato, Dusit and Debbah, Mérouane},
	month = jun,
	year = {2024},
	pages = {117--124},
}

@article{lu_tutorial_2024,
	title = {A {Tutorial} on {Near}-{Field} {XL}-{MIMO} {Communications} {Toward} {6G}},
	volume = {26},
	issn = {1553-877X},
	url = {https://ieeexplore.ieee.org/abstract/document/10496996},
	doi = {10.1109/COMST.2024.3387749},
	number = {4},
	urldate = {2025-08-15},
	journal = {IEEE Communications Surveys \& Tutorials},
	author = {Lu, Haiquan and Zeng, Yong and You, Changsheng and Han, Yu and Zhang, Jiayi and Wang, Zhe and Dong, Zhenjun and Jin, Shi and Wang, Cheng-Xiang and Jiang, Tao and You, Xiaohu and Zhang, Rui},
	year = {2024},
	pages = {2213--2257},
}

@inproceedings{rodrigues_low_2020,
	title = {Low-{Complexity} {Distributed} {XL}-{MIMO} for {Multiuser} {Detection}},
	url = {https://ieeexplore.ieee.org/abstract/document/9145378},
	doi = {10.1109/ICCWorkshops49005.2020.9145378},
	urldate = {2025-08-15},
	booktitle = {2020 {IEEE} {International} {Conference} on {Communications} {Workshops} ({ICC} {Workshops})},
	author = {Rodrigues, Victor Croisfelt and Amiri, Abolfazl and Abrão, Taufik and de Carvalho, Elisabeth and Popovski, Petar},
	month = jun,
	year = {2020},
	note = {ISSN: 2474-9133},
	pages = {1--6},
}

@inproceedings{lu_how_2021,
	title = {How {Does} {Performance} {Scale} with {Antenna} {Number} for {Extremely} {Large}-{Scale} {MIMO}?},
	url = {https://ieeexplore.ieee.org/document/9500972},
	doi = {10.1109/ICC42927.2021.9500972},
	urldate = {2025-08-15},
	booktitle = {{ICC} 2021 - {IEEE} {International} {Conference} on {Communications}},
	author = {Lu, Haiquan and Zeng, Yong},
	month = jun,
	year = {2021},
	note = {ISSN: 1938-1883},
	pages = {1--6},
}

@article{sakhnini_near_field_2022,
	title = {Near-{Field} {Coherent} {Radar} {Sensing} {Using} a {Massive} {MIMO} {Communication} {Testbed}},
	volume = {21},
	issn = {1558-2248},
	url = {https://ieeexplore.ieee.org/abstract/document/9707730},
	doi = {10.1109/TWC.2022.3148035},
	number = {8},
	urldate = {2025-08-15},
	journal = {IEEE Transactions on Wireless Communications},
	author = {Sakhnini, Adham and De Bast, Sibren and Guenach, Mamoun and Bourdoux, André and Sahli, Hichem and Pollin, Sofie},
	month = aug,
	year = {2022},
	pages = {6256--6270},
}

@article{ding_spatial_2024,
	title = {Spatial {Bandwidth} {Asymptotic} {Analysis} for {3D} {Large}-{Scale} {Antenna} {Array} {Communications}},
	volume = {23},
	issn = {1558-2248},
	url = {https://ieeexplore.ieee.org/abstract/document/10213407},
	doi = {10.1109/TWC.2023.3301034},
	number = {4},
	urldate = {2025-08-15},
	journal = {IEEE Transactions on Wireless Communications},
	author = {Ding, Liqin and Zhang, Jiliang and Ström, Erik G.},
	month = apr,
	year = {2024},
	pages = {2638--2652},
}

@article{ding_degrees_2022,
	title = {Degrees of {Freedom} in {3D} {Linear} {Large}-{Scale} {Antenna} {Array} {Communications}—{A} {Spatial} {Bandwidth} {Approach}},
	volume = {40},
	issn = {1558-0008},
	url = {https://ieeexplore.ieee.org/document/9848802},
	doi = {10.1109/JSAC.2022.3196106},
	number = {10},
	urldate = {2025-08-15},
	journal = {IEEE Journal on Selected Areas in Communications},
	author = {Ding, Liqin and Ström, Erik G. and Zhang, Jiliang},
	month = oct,
	year = {2022},
	pages = {2805--2822},
}

@article{zhou_sparse_2025,
	title = {Sparse {Array} {Enabled} {Near}-{Field} {Communications}: {Beam} {Pattern} {Analysis} and {Hybrid} {Beamforming} {Design}},
	issn = {1558-2248},
	shorttitle = {Sparse {Array} {Enabled} {Near}-{Field} {Communications}},
	url = {https://ieeexplore.ieee.org/document/11039147},
	doi = {10.1109/TWC.2025.3578561},
	urldate = {2025-08-15},
	journal = {IEEE Transactions on Wireless Communications},
	author = {Zhou, Cong and You, Changsheng and Zhang, Haodong and Chen, Li and Shi, Shuo},
	year = {2025},
	pages = {1--1},
	}

@article{kosasih_spatial_dof_2025,
	title = {Spatial {Frequencies} and {Degrees} of {Freedom}: {Their} roles in near-field communications},
	volume = {42},
	issn = {1558-0792},
	shorttitle = {Spatial {Frequencies} and {Degrees} of {Freedom}},
	url = {https://ieeexplore.ieee.org/document/10934778},
	doi = {10.1109/MSP.2024.3511922},
	number = {1},
	urldate = {2025-08-15},
	journal = {IEEE Signal Processing Magazine},
	author = {Kosasih, Alva and Demir, Ozlem Tugfe and Kolomvakis, Nikolaos and Bjornson, Emil},
	month = jan,
	year = {2025},
	pages = {33--44}
}

@article{gao_integrated_2023,
	title = {Integrated {Sensing} and {Communication} {With} {mmWave} {Massive} {MIMO}: {A} {Compressed} {Sampling} {Perspective}},
	volume = {22},
	issn = {1558-2248},
	shorttitle = {Integrated {Sensing} and {Communication} {With} {mmWave} {Massive} {MIMO}},
	url = {https://ieeexplore.ieee.org/document/9898900},
	doi = {10.1109/TWC.2022.3206614},
	number = {3},
	urldate = {2025-08-15},
	journal = {IEEE Transactions on Wireless Communications},
	author = {Gao, Zhen and Wan, Ziwei and Zheng, Dezhi and Tan, Shufeng and Masouros, Christos and Ng, Derrick Wing Kwan and Chen, Sheng},
	month = mar,
	year = {2023},
	pages = {1745--1762},
}

@inproceedings{krivosheev_grating_2010,
	title = {Grating lobe suppression in phased arrays composed of identical or similar subarrays},
	url = {https://ieeexplore.ieee.org/abstract/document/5613283},
	doi = {10.1109/ARRAY.2010.5613283},
	urldate = {2025-08-15},
	booktitle = {2010 {IEEE} {International} {Symposium} on {Phased} {Array} {Systems} and {Technology}},
	author = {Krivosheev, Yury V. and Shishlov, Alexandr V.},
	month = oct,
	year = {2010},
	pages = {724--730},
}

@inproceedings{zhuang_coherent_2008,
	title = {Coherent synthesis sparse aperture radar with grating lobes suppressed using frequency {MIMO} technique},
	url = {https://ieeexplore.ieee.org/document/4720971},
	doi = {10.1109/RADAR.2008.4720971},
	urldate = {2025-08-15},
	booktitle = {2008 {IEEE} {Radar} {Conference}},
	author = {Zhuang, Long and Liu, Xingzhao},
	month = may,
	year = {2008},
	note = {ISSN: 2375-5318},
	pages = {1--5},
}

@misc{monnoyer_chirp_2025,
	title = {Chirp-{Based} {Aliasing} {Analysis} of {Arrays} in the {Spherical} {Wavefront} {Regime}},
	url = {http://arxiv.org/abs/2505.05378},
	doi = {10.48550/arXiv.2505.05378},
	urldate = {2025-08-15},
	publisher = {arXiv},
	author = {Monnoyer, Gilles and Defraigne, Laurence and Sambon, Baptiste and Louveaux, Jérôme and Vandendorpe, Luc},
	month = may,
	year = {2025},
	note = {arXiv:2505.05378 [eess]},
}

@article{hafner_calibration_2019,
	title = {On {Calibration} and {Direction} {Finding} with {Uniform} {Circular} {Arrays}},
	volume = {2019},
	copyright = {Copyright © 2019 Stephan Häfner et al.},
	issn = {1687-5877},
	url = {https://onlinelibrary.wiley.com/doi/abs/10.1155/2019/1523469},
	doi = {10.1155/2019/1523469},
	language = {en},
	number = {1},
	urldate = {2025-08-15},
	journal = {International Journal of Antennas and Propagation},
	author = {Häfner, Stephan and Käske, Martin and Thomä, Reiner},
	year = {2019},
	pages = {1523469},
}

@inproceedings{zhang_near_field_2025,
	title = {Near-field {Beam} {Focusing} under {Discrete} {Phase} {Shifters}},
	url = {https://ieeexplore.ieee.org/document/10978226/},
	doi = {10.1109/WCNC61545.2025.10978226},
	urldate = {2025-12-05},
	booktitle = {2025 {IEEE} {Wireless} {Communications} and {Networking} {Conference} ({WCNC})},
	author = {Zhang, Haodong and You, Chang sheng and Zhou, Cong},
	month = mar,
	year = {2025},
	note = {ISSN: 1558-2612},
	pages = {1--6},
}
